\documentclass[12pt]{article}

\usepackage{amsthm,amsmath,amsfonts,amssymb}
\usepackage[authoryear]{natbib}
\usepackage[hypertexnames=false]{hyperref}
\hypersetup{colorlinks,citecolor=blue}
\usepackage{mathtools}

\usepackage{authblk}
\usepackage{fullpage}
\usepackage{mathptmx}
\usepackage[pdftex]{graphicx}
\usepackage[lined,ruled,linesnumbered]{algorithm2e}
\usepackage[export]{adjustbox}
\usepackage{float}
\usepackage{booktabs}
\usepackage{framed}
\usepackage{color}

\providecommand{\keywords}[1]
{
  \small	
  \textbf{\textit{Keywords---}} #1
}

\usepackage{xspace}
\usepackage{stmaryrd}

\newcommand{\gof}{GoF\xspace}
\newcommand{\lof}{LOF\xspace}
\newcommand{\knn}{kNN\xspace}

\newcommand{\hpc}{HPC\xspace}
\newcommand{\PODs}{PODs\xspace}
\newcommand{\POD}{POD\xspace}

\newcommand{\given}[2]{\left.#1\mathrel{}\middle|\mathrel{}#2\right.} 
\newcommand{\Proba}{\mathbb{P}}
\newcommand{\Esp}{\mathbb{E}}

\newcommand{\Var}{\mathbb{V}\text{ar}}

\newcommand{\intervalle}[4]{\mathopen{#1}#2\mathclose{}\mathpunct{},#3\mathclose{#4}}
\newcommand{\intint}[2]{\intervalle\llbracket{#1}{#2}\rrbracket}
\newcommand{\set}[1]{\left\{#1\right\}}
\newcommand{\ind}[1]{\mathbb{I}\left\{{#1}\right\}}
\DeclareMathOperator*{\argmin}{argmin}
\DeclareMathOperator{\KL}{KL}
\newcommand{\card}[1]{\left| #1 \right|}
\DeclareMathOperator*{\logit}{logit}
\DeclareMathOperator*{\logitinv}{logit^{-1}}
\DeclareMathOperator*{\round}{round}

\newcommand{\model}{m}
\newcommand{\paramSetModel}[1]{\Theta_{#1}}
\newcommand{\paramSet}{\paramSetModel{\model}}
\newcommand{\param}{\boldsymbol{\theta}}
\newcommand{\priorModel}[1]{\pi_{#1}}
\DeclareMathOperator{\prior}{\priorModel{\model}}
\newcommand{\priorPredModel}[1]{p_{#1}}
\DeclareMathOperator{\priorPred}{\priorPredModel{\model}}
\newcommand{\postPredModel}[3]{p_{#3}\left(\given{#1}{#2}\right)}
\newcommand{\postPred}[2]{\postPredModel{#1}{#2}{\model}}
\newcommand{\refTableModel}[1]{\mathbf{D}_{#1}}
\newcommand{\refTable}{\refTableModel{\model}}
\newcommand{\refTableHat}{\hat{\mathbf{D}}_{\model}}
\newcommand{\refTableHatModel}[1]{\hat{\mathbf{D}}_{#1}}
\newcommand{\like}[2]{f_{\model}\left(\given{#1}{#2}\right)}
\newcommand{\obsSet}{\mathcal{Y}}
\newcommand{\y}{\mathbf{y}}
\renewcommand{\u}{\mathbf{u}}
\newcommand{\yobs}{\y_{\text{obs}}}
\newcommand{\yrep}{\y_{\text{rep}}}
\newcommand{\ynew}{\y_{\text{new}}}
\newcommand{\rawObs}{\mathbf{z}}
\DeclareMathOperator{\obsDistr}{F}

\newcommand{\iref}{i}
\newcommand{\nref}{N_{\text{ref}}}
\newcommand{\icalib}{j}
\newcommand{\ncalib}{N_{\text{calib}}}
\newcommand{\npost}{N_{\text{post}}}
\newcommand{\ntest}{N_{\text{test}}}
\newcommand{\itest}{k}

\newcommand{\scoreRefTable}[2]{T\left(#1; #2\right)}
\newcommand{\score}[1]{\scoreRefTable{#1}{\refTable}}
\newcommand{\scoreModel}[2]{\scoreRefTable{#1}{\refTableModel{#2}}}

\begin{document}

\title{Goodness of Fit for Bayesian Generative Models with Applications in Population Genetics\thanks{All authors contributed equally to this work.}}

\date{\vspace*{-0.25cm} January 28, 2025}

\author[1]{Guillaume Le Mailloux}

\author[2]{Paul Bastide}

\author[1]{Jean-Michel Marin\thanks{Corresponding author: jean-michel.marin@umontpellier.fr}}

\author[3]{Arnaud Estoup}

\affil[1]{IMAG, Univ. Montpellier, CNRS, Montpellier, France}

\affil[2]{MAP5, Univ. Paris Cit\'e, CNRS, Paris, France}

\affil[3]{CBGP, INRAE, CIRAD, IRD, Montpellier SupAgro, Univ. Montpellier, Montpellier, France}

\maketitle

\vspace*{-0.95cm}

\begin{abstract} 
In population genetics and other application fields, models with intractable likelihood are common.
Approximate Bayesian Computation (ABC) or more generally Simulation-Based Inference (SBI) methods work by simulating instrumental data sets from the models under study and comparing them with the observed data set $\yobs$, using advanced machine learning tools for tasks such as model selection and parameter inference.
The present work focuses on model criticism, and more specifically on Goodness of fit (\gof) tests, for intractable likelihood models.
We introduce two new \gof tests: the pre-inference \gof tests whether $\yobs$ is distributed from the prior predictive distribution,
while the post-inference \gof tests whether there is a parameter value such that $\yobs$ is distributed from the likelihood with that value. The pre-inference test can be used to prune a large set of models using a limited amount of simulations, while the post-inference test is used to assess the fit of a selected model.
Both tests are based on the local outlier factor (\lof, \citealp{Breunig2000}).
This indicator was initially defined for outlier and novelty detection.
It is able to quantify local density deviations, capturing subtleties that a more traditional \knn-based approach may miss.
We evaluated the performance of our two \gof tests on simulated datasets from three different model settings of varying complexity.
We then illustrate the utility of these approaches on a dataset of single nucleotide polymorphism (SNP) markers for the evaluation of complex evolutionary scenarios of modern human populations.
Our dual-test \gof approach highlights the flexibility of our method: the pre-inference \gof test provides insight into model validity from a Bayesian perspective, while the post-inference test provides a more general and traditional view of assessing goodness of fit.
\end{abstract}


\section{Introduction}

In numerous applications, the direct calculation of likelihood functions is unfeasible, thereby posing considerable challenges for inference. Two common cases serve to exemplify this issue. The first case pertains to latent variable models, wherein the likelihood function necessitates the evaluation of an intractable integral: $f(\y|\param)=\int f(\y,\u|\param) \mu(\text{d}\u)$. This is particularly challenging due to the hidden structure of the model. The second case pertains to likelihoods that are expressed as follows: $f(\y|\param)=g(\y,\param)/Z(\param)$ and $Z(\param)$ intractable. Sophisticated inferential techniques are required in both cases. In this paper, we will focus on cases where it is possible to simulate from the likelihood, which is typically more straightforward in the case of a latent variable model.

\subsection{Statistical Inference with Intractable Likelihood}

Traditional solutions to these scenarios include expectation-maximisation (EM) algorithms, Gibbs sampling, pseudo-marginal Markov chain Monte Carlo (MCMC) methods and variational approximations.
However, these methods rely on the explicit computation of
the likelihood $f(\y,\u|\param)$. In Population Genetics and many other application fields,
complex models with intractable likelihoods are common.
Approximate Bayesian Computation (ABC; \citealp{Marin2012}) has emerged as a significant methodology for inference when the likelihood is intractable but data generation from the model is feasible. ABC, also known as Simulation-Based Inference (SBI) in the context of machine learning \citep{Cranmer2020}, circumvents the direct calculation of the likelihood by comparing simulated pseudo-datasets to the observed dataset, thereby facilitating inference without the explicit computation of the likelihood.

Since the seminal contributions of \cite{TavareBGD97} and \cite{Pritchardx99}, ABC/SBI has undergone substantial developments. Most popular and effective strategies involve applying Machine Learning tools to a training set generated by a Bayesian generative model (the reference table). 
We review below some of the recent developments in the field,
with a focus on applications to Population Genetics,
and refer to \citet{Marin2012} and \citet{Cranmer2020} for comprehensive reviews,
and to \citet{LueckmannX21} for an extensive benchmark of simulation-based inference methods.
Recently,
\cite{PapamakariosM16} introduced a method for fast likelihood-free inference using Bayesian Conditional Density Estimation with Mixture Density Networks, which has been shown to be effective in estimating the posterior distribution, particularly in the context of complex models. \citet{Pudlox16} put forth an approach for dependable ABC model selection utilizing classification random forests, markedly enhancing model choice accuracy by capitalizing on machine learning techniques in a bioinformatics context. Moreover, \cite{SheehanS16} employed deep learning techniques, specifically unsupervised pre-training via autoencoders, in the context of population genetic inference, thereby illustrating the potential of neural networks for the extraction of informative features in the context of ABC. Building on these advancements, \citet{Raynal2018} developed ABC random forests for Bayesian parameter inference, employing regression random forests to estimate parameters within the ABC framework. 

In a more recent development, \citet{PapamakariosSM19} introduced Sequential Neural Likelihood Estimation, utilizing autoregressive flows for neural density estimation of the likelihood. This enables fast and accurate inference with normalizing flows, including  Masked Autoregressive Flows \citep{PapamakariosPM17}. \citet{GreenbergNM19} introduced a method called automatic posterior transformation for likelihood-free inference, which also uses neural density estimation. This method is different from neural likelihood estimation because it uses normalizing flows to directly estimate the posterior. This method is part of the class of Sequential Neural Posterior Estimation, like \cite{LueckmannGBONM17}. Another notable contribution was made by \citet{Hermansx20}, who introduced Likelihood-free MCMC with approximate likelihood ratios through Sequential Neural Ratio Estimation. This method employs neural networks to estimate likelihood ratios, thereby providing a novel approach to efficient likelihood-free inference. 

\subsection{Model Criticism}

The present work focuses on model criticism, which is essential for the assessment of whether an observed dataset, $\yobs$, is likely to have been generated by a given model.
In the context of likelihood-free inference, \cite{RatamnnAWR09} discuss the benefit of incorporating model diagnostics within an ABC framework, and demonstrate how this method diagnoses model mismatch and guides model refinement. \cite{Wilkinson13} shown that ABC gives exact results under the assumption of model error. More recently, \cite{FrazierRR20} explore the effects of model misspecification in ABC, showing that different ABC methods produce varying results when the simulation model deviates from the true data-generating process. It finds that accept-reject ABC concentrates on a pseudotrue value but lacks valid frequentist coverage, while local regression ABC targets a different pseudotrue value. The study also proposes diagnostic methods to detect model misspecification, supported by theoretical analysis and practical examples. Finally,  \cite{Ward2022} discuss the issue of model misspecification in Neural Posterior Estimation, highlighting how traditional approaches often assume that simulators are perfectly specified representations of reality. It is demonstrated that such assumptions result in unreliable inferences when discrepancies exist between simulated and observed data. To address this issue, they propose Robust Neural Posterior Estimation, a methodology that explicitly models these discrepancies, thereby enabling both model criticism and robust inference. This approach ultimately mitigates the risks associated with misspecification and produces more reliable posterior estimates. 

The present work focuses on Bayesian model checking. 
This type of analysis is important for model validation. It can also be used for model pruning when the number of candidates to be compared is excessive, especially in the context where data simulation is expensive. Bayesian model checking entails the assessment of the degree of fit between a Bayesian model and the observed data. As discussed by \cite{Box1980}, the prior predictive  check assesses model assumptions without consideration of the data. It involves a comparison of observed data with data simulated from the prior predictive distribution (the likelihood is integrated over the prior, also called the evidence), with the aim of identifying major discrepancies between the simulated data and the observed dataset.
In the posterior predictive check, as outlined by \citet{Guttman1967}, \citet{Rubin1984}, and \citet{Gelman1996}, observed data is compared to data simulated from the posterior predictive distribution (the likelihood is integrated over the posterior distribution). More recent approaches, such as the holdout method proposed by \citet{Moran2023}, utilize a portion of the data as a validation set to evaluate model performance on unseen data.
\citet{Lemaire2016} presented a test for goodness of fit in the prior predictive check framework, that makes use of k-nearest neighbour (k-NN) density estimation. 
This approach circumvents the necessity for explicit likelihood functions by estimating the density of  observed and simulated data to measure misfit.
In the present work, we focus on (i) improving the model misfit measure by drawing
from the novelty detection literature and (ii) building
a correctly calibrated post-inference goodness of fit test using the holdout approach.

\subsection{Goodness of Fit as a Novelty Detection Problem}
In this paper, we link the question of likelihood-free goodness of fit to the field of novelty and outlier detection in order to improve on the results obtained by \citet{Lemaire2016}.  Novelty/outlier detection is used to identify anomalous or infrequent data points within a given dataset. Novelty detection is the process of identifying previously unseen patterns in new data, whereas outlier detection is the method of locating data points that deviate significantly from the majority in an existing dataset. These techniques are of particular importance in fields such as anomaly detection, fraud detection, and fault monitoring. \citet{Goldstein2016} discuss various unsupervised outlier detection techniques, emphasizing their scalability and adaptability to high-dimensional data. \citet{Han2022} explore more recent advancements in anomaly detection methods, particularly in the context of machine learning.

One frequently employed method for novelty/outlier detection is the Local Outlier Factor (\lof; \cite{Breunig2000}. The \lof identifies anomalous data points based on their local density relative to their neighbors, thereby enabling the effective detection of outliers in detasets where clusters of varying point densities are present. This method assesses the degree of isolation of a given point in comparison to its surrounding context, rendering it particularly useful for datasets exhibiting diverse density structures, where more global methods may prove ineffective.
Figure~\ref{fig:example_dep_indep_lof} illustrates this behavior.
It shows the first principal component axes of summary statistics of 
data points simulated according to a simple Population Genetics model
(namely, the scenario 2 of the so-called Dep-Indep setting detailed below).
The simulated distribution of points (blue circles) has an heterogeneous density,
with a densely populated region at the center, and a sparser region
on the lower left corner of the plot.
The first test point (black triangle) is located at the border of the dense
region, so that its distance to its neighbors as well as its local 
density are typical of the simulated points around.
The prior \gof test described in this article does not reject the
null hypothesis, whether using the \knn or \lof based score.
The second test point (black square) is isolated,
so that its local density is lower than that of its closest simulated points: in this case the prior \gof test based on the \lof score identifies this local density discrepancy and rejects the null hypothesis. On the other hand, the \gof test based on the \knn score does not reject the null hypothesis in this case again, because the raw distance of the second test point to its neighbors is not atypical among the simulated points at the level of the dataset as a whole due to the presence of sparse regions, especially in the bottom left area.
In what follows, we will show that this behavior extends consistently to large and complex models, which leads to an improvement in the power of the test based on the \lof score.

\subsection{Statistical Inference in Population Genetics}

A significant number of ABC/SBI methods have been developed by population geneticists \citep{TavareBGD97,Pritchardx99,Beaumont2002}.
These methods are particularly valuable in this area due to the high complexity of demo-genetic models and the often intractable or implicit nature of the likelihood functions associated with these models. ABC/SBI permit researchers to test evolutionary hypotheses through model choice and estimate parameters even when traditional likelihood-based approaches are infeasible. For example, applications include the inference of demographic history, selection, migration rates, and recombination patterns. By focusing on summary statistics that capture essential key patterns in the data, ABC/SBI methods allow for scalable and flexible analyses, thus making them essential tools in modern population genomics \citep{Beaumont2010,Beaumont19}. 

To illustrate how powerful our likelihood-free goodness of fit test can be in the field, we worked on a case study (relying both on simulated and real genomic datsets) involving complex evolutionary scenarios among human populations \cite{Pudlox16}. We started with six hypothetical evolutionary scenarios to model four modern human populations. We chose two of these scenarios, which are quite similar but really important in terms of real human SNP data, to look at in more detail. This setup is a great way to test how well modern computational approaches can distinguish between different evolutionary paths, which are often complex and related, and to assess to what extent do the observed data fit with the scenario considered to be the best among those proposed.

\subsection{Outline}

The paper is structured as follows. In section \ref{sec:methods}, we outline our method, detailing the computational framework.  In section \ref{sec:simulation}, we test our method through simulation studies, applying it to a range of hypothetical evolutionary scenarios to evaluate its accuracy and robustness. Section \ref{sec:human_appli} then illustrates our approach using a real SNP dataset from modern human populations, providing empirical evidence of its practical utility.  A discussion concludes the paper.

\begin{figure}[H]
    \centering
    \includegraphics[]{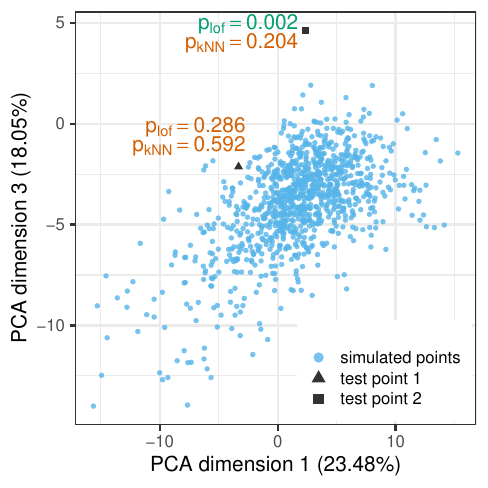}
    \caption{
    \textbf{Illustrative example of different or similar conclusions when using tests based on the \lof or the \knn score from the Dep-Indep setting}
    (see Section~\ref{sec:depindep_setting}).
    Axes 1 and 3 from a PCA on the summary statistics are represented.
    Simulated points (blue) are from scenario 2,
    while test points (black) are from scenario 1 of the Dep-Indep setting.
    The \lof score is computed with the ``max-\lof'' heuristic with $k$ varying between $5$ and $20$,
    while the \knn uses $k = 1$ (see Section 2.2 Outlier Score).
    The square point (top) is correctly rejected by the \lof, but not rejected by the \knn.
    It is far from the other points locally, so that it has a hight \lof score,
    but its distance to nearby points is not an outlier when compared
    to typical distances in the whole dataset (particularly in the bottom left region),
    so its \knn score is not outlying.
    The triangle point is not rejected by either the \lof or \knn scores:
    its distance from other points does not differ from that of its neighbors.
    }
    \label{fig:example_dep_indep_lof}
\end{figure}

\section{Methods}\label{sec:methods}

\subsection{Model}

Let $\paramSet$ be the space of possible parameter values for a given model $\model$,
with a prior distribution $\prior$ and likelihood function
$\like{\y}{\param}$ that gives a distribution on the observation space $\obsSet$
given a vector of parameters $\param \in \paramSet$.
We are typically interested in cases where this likelihood is intractable,
and we assume that the model is approximated by simulated data recorded in a reference table $\refTable$:
\[
\refTable = \set{(\param_\iref, \y_\iref), \iref \in \intint{1}{\nref}},
\text{ with } \param_\iref\underset{\text{iid}}{\sim}\prior 
\text{ and } \y_\iref\underset{\text{indep.}}{\sim}\like{\cdot}{\param_\iref}. 
\]

Typically, the observation space $\obsSet$ has a very high dimension,
and a set of summary statistics is used to reduce the data,
so that the reference tables has summarized observations $\eta(\yobs)$
instead of full observations. 
Recent methods use machine learning techniques to learn good summary statistics
directly from the data \citep{Fearnhead2012,Forbes2022}, but this remains a
difficult task in a general setting.
However, in some specific fields such as population genetics, 
informative summary statistics based on expert knowledge are available
\citep{Pudlox16,Raynal2018}.
In all the following, we assume that a set of such summary statistics is
already known for the problem at hand.
With a slight abuse of notation, we will write $\yobs$ for $\eta(\yobs)$,
so that all observations are in the summary statistics space.

\subsection{Outlier Score}
For any observation $\y \in \obsSet$, we define an outlier score
$\score{\y}$ that should be designed to be small for points
$\y$ that are ``inside'' the distribution represented by $\refTable$,
and large for points that are ``outside'' of this distribution.

\citet{Lemaire2016} use the mean \knn score, that is defined as the
mean distance to the $k$ nearest neighbors of a point.
More specifically, for $\y \in \obsSet$, we denote by
$N_k(\y;\refTable)$ the set of the $k$ nearest neighbors of $\y$ in the
reference table $\refTable$ (excluding $\y$),
and set:
\begin{equation}\label{eq:def_knn}
\mbox{knn}(\y, \refTable) = 
\frac{1}{k} \sum_{\y'\in N_k(\y;\refTable)} d(\y, \y'),
\end{equation}
with $d$ the Euclidean distance between $\y$ and $\y'$.
Note that in \citet{Lemaire2016}, this score is not explicitly linked 
to the \knn mean distance.
They also set $k$ so that a given fraction $\epsilon$ of the
reference table is used in the computation of the score (typically, $\epsilon = 1\%$).
In the following, we use our simulation study to give some guidelines to select 
the value of $k$ (see Section~\ref{sec:score_computation_choice_k}).

In the present work, inspired from the literature on novelty detection,
we use the \lof score \citep{Breunig2000}.
It relies on the k-local reachable density in $\y$ with respect to the reference table $\refTable$:

\begin{equation}\label{eq:def_lrd}
\mbox{lrd}_k(\y;\refTable) = 
\left(\frac{1}{k}\sum_{\y'\in N_k(\y;\refTable)} \text{reach-dist}_k(\y,\y') \right)^{-1},
\end{equation}
where $\text{reach-dist}_k(\y,\y')=\max\{d(\y,\y');\ \text{k-dist}(\y';\refTable)\}$,
with  $\text{k-dist}(\y';\refTable)$ the $k^{th}$ nearest distance 
between $\y'$ and points in $\refTable$.
Note that if we were to replace the $\text{reach-dist}$ by the simple euclidean distance 
in the definition above, we would have $\mbox{lrd}_k(\y;\refTable) = \mbox{knn}(\y, \refTable)^{-1}$.
\citet{Breunig2000} use the $\text{reach-dist}$ to stabilize the $\text{lrd}$.
To take into account distributions with heterogeneous densities,
where some regions of the space are much more dense that others
(see Fig.\ \ref{fig:example_dep_indep_lof} for an illustration)
the \lof statistic is defined as a ratio:

\begin{equation}\label{eq:def_lof}
\text{\lof}_k(\y;\refTable) = 
\frac{\frac{1}{k}\sum_{\y'\in N_k(\y;\refTable)} \text{lrd}_k(\y';\refTable)}{\text{lrd}_k(\y;\refTable)}.
\end{equation}

This statistic is fully local and can be easily interpreted: 
for instance, $\text{\lof}_k(\y;\refTable) \approx 3$ means that the density in $\y$ is 
three times smaller than the average density of its neighborhood.
As in the \knn case, the value of $k$ must be chosen by the practitioner.
\cite{Breunig2000} suggest to define a ``max-\lof'' statistic: 
$\text{max-\lof}(\y;\refTable) = \underset{k\in I}{\text{max}}\ \text{\lof}_k(\y;\refTable)$ 
where $I$ is an interval of integers between, \textit{e.g}., 5 and 20.

\subsection{Pre-Inference Prior \gof} \label{sec:prior_gof}

The goal of the pre-inference goodness of fit test (hereafter pre-inference \gof) is to check whether
an observed data $\yobs \in \obsSet$ is consistent with the prior predictive distribution
of a given model $\model$. Let $\obsDistr$ be the true distribution function
that generated the data. The null hypothesis associated with the prior-\gof test
can hence be written as:
\begin{equation}\label{eq:prior_GoF_hyp}
H_0^{\text{prior}} : \obsDistr(\cdot) = \priorPred(\cdot) \coloneqq \int_{\paramSet}\like{\cdot}{\param}\prior(d\param).
\end{equation}
Following \citet{Lemaire2016}, we then define the p-value associated with this test
by the probability of the observed score $\score{\yobs}$ to be larger than the
score $\score{\yrep}$ of replicates $\yrep \sim \priorPred$
randomly drawn from the prior predictive:
\begin{equation}\label{eq:prior_GoF_def}
P^{\text{prior}}(\yobs) = \Proba_{\yrep \sim \priorPred}\left[\score{\yrep} > \score{\yobs} \right].
\end{equation}
Note that by definition, this p-value is well calibrated:
when $H_0$ is true, $\yobs \sim \priorPred$, and the p-value
is uniformly distributed between $0$ and $1$.
To estimate this p-value, we use $\ncalib$ calibration points drawn from the 
prior predictive $\y_{\icalib} \underset{\text{iid}}{\sim} \priorPred$,
and compare the observed score with the score of the replicates:
\begin{equation}\label{eq:prior_GoF_emp}
\hat{P}^{\text{prior}}(\yobs) = 
\frac{1}{\ncalib} \sum_{\icalib = 1}^{\ncalib} \ind{\score{\y_{\icalib}} > \score{\yobs}}.
\end{equation}

\subsection{Post-Inference Holdout \gof}\label{sec:post_gof}

The prior \gof as described above is a frequentist test, based on an empirical p-value,
of a Bayesian null hypothesis that the data was distributed according to the prior predictive distribution.
Taking one step further in the frequentist framework, our goal with the post-inference holdout \gof (hereafter post-inference \gof) is to test whether the true distribution $\obsDistr$ is equal to the maximized likelihood of the model:

\begin{equation}\label{eq:holdout_GoF_hyp}
H_0^{\text{holdout}} : \obsDistr(\cdot) = \like{\cdot}{\param_{\model}^{*}}
\quad
\text{with}
\quad
\param_{\model}^{*} = \argmin_{\param\in\paramSet}\KL\left[\obsDistr(\cdot) \;\vert\vert\; \like{\cdot}{\param}\right],
\end{equation}
with $\KL$ the Kullback-Leibler divergence.
From this null hypothesis and as in the previous section, we can define the following theoretical p-value:
\[
P^{\text{freq}}(\yobs) = 
\Proba_{\yrep \sim \like{\cdot}{\param_{\model}^{*}}}
\left[\scoreRefTable{\yrep}{\refTable^{*}} > \scoreRefTable{\yobs}{\refTable^{*}} \right],
\]
where $\refTable^{*}$ would be a reference table drawn from $\like{\cdot}{\param_{\model}^{*}}$
instead of $\priorPred(\cdot)$.
As in the pre-inference-\gof test, this p-value is, by design, uniformly distributed under the null
hypothesis.
Note that we could use any reference table in the score $T$ used above, including 
the original $\refTable$ drawn from the prior predictive, and this p-value would
still be well calibrated.
However, using the (theoretical) $\refTable^{*}$ reference table should increase the
discrimination power of the test, as, by design, we expect $\scoreRefTable{\yobs}{\refTable^{*}}$
to be large when $\yobs$ is not drawn according to $H_0$, and small otherwise.

Unfortunately, as opposed to the prior-\gof test, where generating data from the null hypothesis $\priorPred(\cdot)$
was straightforward, generating data according to $\like{\cdot}{\param_{\model}^{*}}$
is not feasible, as this quantity is unknown.
The strategy applied by the holdout predictive check is to approximate this quantity 
using the posterior distribution learned from the data. Indeed, under regularity conditions, the posterior distribution asymptotically concentrates around the maximum likelihood estimate as the sample size increases, following the Bernstein-von Mises theorem \citep{vanderVaart1998}.
In order to avoid any double use of the data, \citet{Moran2023} assume that part of the observations
was hold-out, so that we have access to $\ynew \sim \obsDistr(\cdot)$,
an independent observation drawn from the true distribution,
and that we draw replicates from the predictive posterior distribution 
$\postPred{\cdot}{\yobs} = \int_{\paramSet}\like{\cdot}{\param}p(\given{d\param}{\yobs})$
learned from $\yobs$:
\begin{equation}\label{eq:holdout_GoF_def}
P^{\text{holdout}}(\yobs, \ynew) = 
\Proba_{\yrep \sim \postPred{\cdot}{\yobs}} 
\left[
\scoreRefTable{\yrep}{\refTable^{\lvert\yobs}} > \scoreRefTable{\ynew}{\refTable^{\lvert\yobs}}
\mathrel{}\middle|\mathrel{} \yobs, \ynew
\right],
\end{equation}
where $\refTable^{\lvert\yobs}$ would be a reference table drawn from $\postPred{\cdot}{\yobs}$.
Here, the draws from the replicates no longer follow the same
distribution as $\yobs$ under $H_0$, so this p-value is no longer
exactly calibrated. However, under some common assumption, \citet{Moran2023}
showed that the p-value is asymptotically calibrated when the number of observations grows.

In the context of generative models, we typically do not have access to
$\postPred{\cdot}{\yobs}$, or even to a good approximation of it.
However, ABC techniques can allow us to get observations that are approximately
drawn from $\postPred{\cdot}{\yobs}$.
Assume that we have access to $\npost = \ncalib' + \nref'$ such draws, that
we split between $\ncalib'$ calibration points $\y_1, \cdots, \y_{\ncalib'}$,
and $\nref'$ reference points stored in a new approximated posterior reference table
$\refTableHat^{\lvert\yobs}$.
Then the target p-value can be approximated by:
\begin{equation}\label{eq:holdout_GoF_emp}
\hat{P}^{\text{holdout}}(\yobs, \ynew) = 
\frac{1}{\ncalib'} \sum_{\icalib = 1}^{\ncalib'} 
\ind{\scoreRefTable{\y_\icalib}{\refTableHat^{\lvert\yobs}} >
\scoreRefTable{\ynew}{\refTableHat^{\lvert\yobs}}}.
\end{equation}
This quantity can be computed from the observed dataset, as long as we assume that
we can split the data into two independent observations $(\yobs,\ynew)$,
and that we can draw points from the approximated posterior.
This last step is crucial, as the quality of the estimation of the posterior
will control the quality of the approximation of the maximum likelihood that
we are aiming at, and hence will impact the calibration of our p-value.
In an ABC framework, a number of techniques have been proposed for posterior estimation,
including methods based on a simple rejection, a local linear regression \citep{Beaumont2002}, 
a regression with a Ridge regulation algorithm, or a neural network regression \citep{Blum2010}; we refer to
\citet{Beaumont2010,Marin2012} for reviews of available techniques.

\section{Simulation Study}\label{sec:simulation}

\subsection{Material and Methods}

We considered three datasets, described here from the most simple to the most complex,
with the last two taken from established models in population genetics.
In each case, we defined a model $m_0$ that was used to simulate the reference table,
and a model $m_1$ that we used to simulate ``pseudo-observed datasets'' (\PODs),
i.e., datasets that mimic observations coming from an empirical dataset.
Our goal is to apply our goodness of fit procedures (prior-\gof and post-\gof) on such simulated datasets to assess whether model $m_0$ under scrutiny is good enough to describe each \POD from $m_1$.

\subsubsection{Toy ``Laplace-Gaussian'' Model}

We first considered a ``toy'' dataset, with $m_0$ (Laplace) and $m_1$ (Gaussian) defined as follow:
\begin{equation*}
   \text{$m_0$: }
      \left\{
   \begin{aligned}
   \param &= (\mu, \sigma) \\
   \mu &\sim \mathcal{U}(-5,5); \\
   \sigma &\sim \mathcal{U}(1,4);\\
    \rawObs_i &\overset{iid}{\sim} \mathcal{L}(\mu,\sigma/\sqrt{2}) \mbox{ for } 1 \leq i \leq d;\\
    \yobs &= \text{L-moments}(\rawObs; m);
    \end{aligned}
    \right.
   \quad
   \text{$m_1$: }
   \left\{
   \begin{aligned}
   \param &= (\mu, \sigma) \\
    \mu &\sim \mathcal{U}(-5,5); \\
   \sigma &\sim \mathcal{U}(1,4);\\
    \rawObs_i &\overset{iid}{\sim} \mathcal{N}(\mu,\sigma^2) \mbox{ for } 1 \leq i \leq d;\\
    \yobs &= \text{L-moments}(\rawObs; m).
    \end{aligned}
    \right.
\label{eq:gaussianlaplace_models}
\end{equation*}
for each raw data $\rawObs_i$, $1 \leq i \leq d$, both models have conditional mean $\mu$ and standard deviations $\sigma$ as variable parameters draw into uniform prior distributions with bounds (-5,5) and (1,4), respectively.
As the Laplace model has heavier tails than the Gaussian model, rejecting
the Laplace model with a Gaussian observation can be difficult, which is why
we selected this configuration.
We simulated from these models with the \texttt{R} statistical programming language,
using the function \texttt{rlaplace} from the package \texttt{VGAM} \citep{VGAM} for Laplace simulation.
To mimic the standard case where summary statistics are used,
the simulated vectors $\rawObs$ of dimension $d=350$ were not observed directly,
but instead summarized using the first $m = 20$
sample L-moments ratios \citep{Hosking1990} using function 
\texttt{salmu} from package \texttt{lmom} \citep{Hosking2023}.

\subsubsection{Simple ``Dep-Indep'' population genetics model with four populations} \label{sec:depindep_setting}

\begin{figure}[H]
   \begin{center}
     \includegraphics[height=5cm,width=0.6\textwidth]{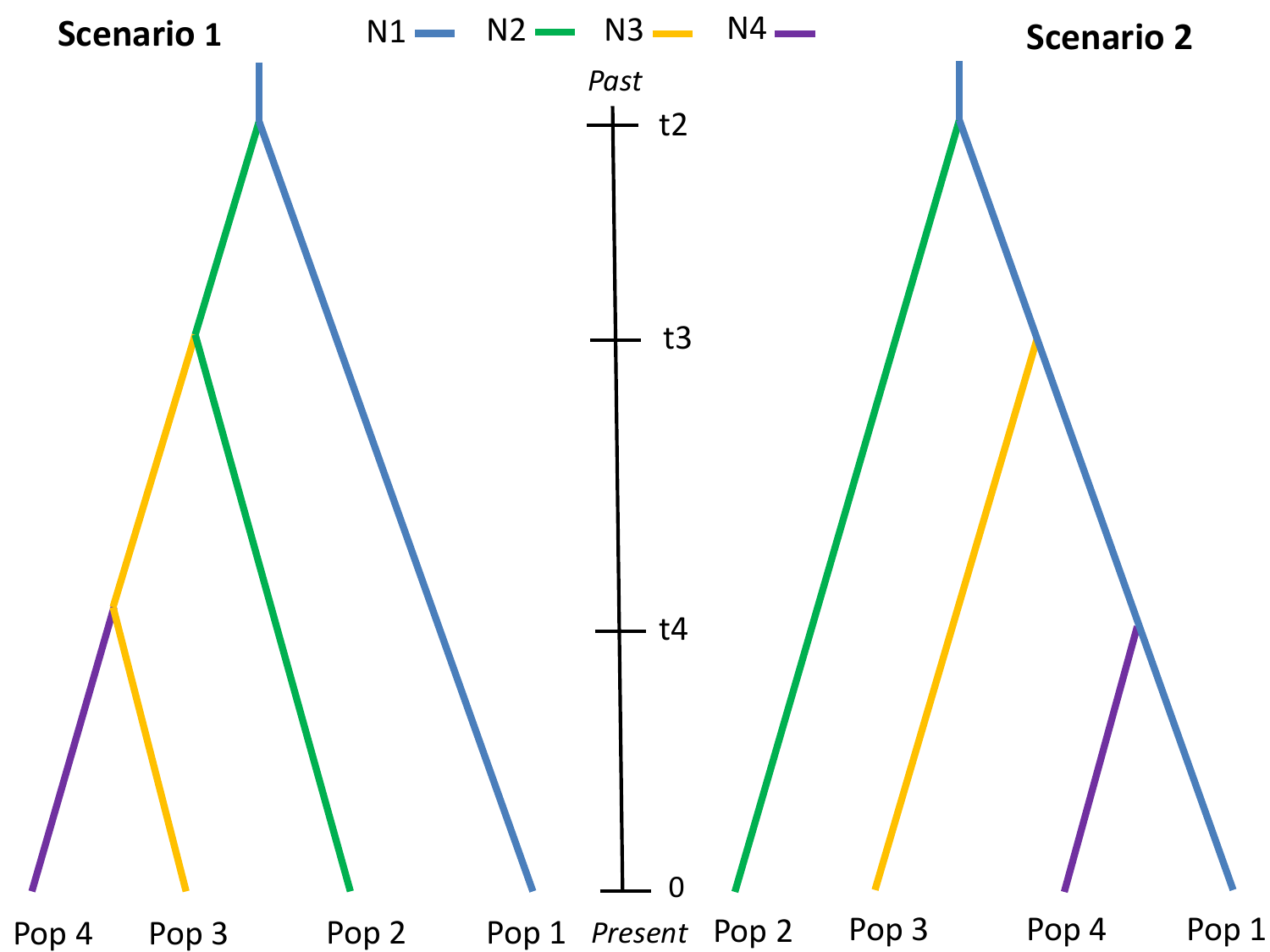}
   \end{center}
   \caption{
   \textbf{Illustrations of the two Dep-Indep evolutionary scenarios}
   considered in the population genetics example of Section~\ref{sec:depindep_setting}.
   In Scenario 1 (``dep''), populations diverged serially from each other,
   population 1 being the ancestral population.
   In Scenario 2 (``indep''), the populations 3, 4 and 4 diverged independently from the ancestral population 1.}
   \label{fig:DepIndep4pop}
 \end{figure}

We considered a case study with two evolutionary scenarios of four populations
using 1000 single nucleotide polymorphism (SNP) genetic markers generated following the ``Hudson'' criterion
\citep{Cornuet2014}
and genotyped for 10 individuals per population.
These two scenarios are depicted in Fig.~\ref{fig:DepIndep4pop}.
In scenario 1 (``dep''), populations diverged serially from each other, population 1 being the ancestral population
(pop 1 $\rightarrow$ pop 2 $\rightarrow$ pop 3 $\rightarrow$ pop 4).
In scenario 2 (``indep''), populations 2, 3 and 4 diverged independently from the ancestral population 1 
(pop 1 $\rightarrow$ pop 2, pop 1 $\rightarrow$ pop 3, and pop 1 $\rightarrow$ pop 4).

These scenarios are often opposed when analyzing invasion or colonization histories,
with scenario 1 corresponding to the case of a cascade of secondary invasion (colonization)
events after a primary invasion (colonization) event 
and scenario 2 corresponding to the case of multiple independent invasion (colonization) events
from a single source population. 
In both models, the parameters are
$\param = (N1, N2, N3, N4, t2, t3, t4)$,
with $Ni$, $1 \leq i \leq 4$,  the effective population size of population $i$ (in number of diploid individuals),
and $ti$, $2 \leq i \leq 4$, the split time (in number of generations since present) between populations $i$ and population $i-1$ for scenario 1,
or between population $i$ and population $1$ for scenario 2.
We used the two following Bayesian models:
\begin{equation*}
   \text{$m_0$ (Sc. 2): }
      \left\{
   \begin{aligned}
    Ni & \sim \mathcal{U}(10^3,10^4), ~ 1 \leq i \leq 4;\\
    t4 &\sim \mathcal{U}(1,60);\\
    t3 &\sim \mathcal{U}(61,120);\\
    t2 &\sim \mathcal{U}(121,180);\\
    \rawObs &\sim \mbox{Scenario\ 2}(\param);\\
    \yobs &= \text{SumStat}(\rawObs; m);
    \end{aligned}
    \right.
   \quad
   \text{$m_1$ (Sc. 1): }
   \left\{
   \begin{aligned}
    Ni & \sim \mathcal{U}(10^3,10^4), ~ 1 \leq i \leq 4;\\
    t4 &\sim \mathcal{U}(1,60);\\
    t3 &\sim \mathcal{U}(61,120);\\
    t2 &\sim \mathcal{U}(121,180);\\
    \rawObs &\sim \mbox{Scenario\ 1}(\param);\\
    \yobs &= \text{SumStat}(\rawObs; m).
    \end{aligned}
    \right.
\label{eq:depindep_models}
\end{equation*}
We simulated SNP datasets from each scenario with the diyabc simulation program (\citealp{Collin2021}\footnote{Available at: \url{https://github.com/diyabc/diyabc}.}; release v1.1.51) associated to the package DIYABC-RF v1.0 (\citealp{Collin2021}\footnote{Available at: \url{https://github.com/diyabc/diyabc.github.io}.}).
The observed and simulated SNP datasets were summarized using a total of $m = 130$ summary statistics
describing genetic variation within populations 
(\textit{e.g}., proportion of monomorphic loci, heterozygosity, population-specific FST)
and between pair, triplet or quadruplet of populations 
(\textit{e.g}., Nei’s distance, Fst-related statistics, Patterson’s allele-sharing f-statistics,
coefficients of admixture)
to describe genetic variation among various population combinations (see  \citealp{Collin2021} for details).

\subsubsection{More complex models inspired by evolutionary scenarios for human populations}\label{sec:sim_human_like}

For this third simulation-based case study, we considered two complex evolutionary scenarios
derived from the study of four populations of modern humans initially comprising six evolutionary
scenarios and used previously in other methodological studies (e.g., \citealp{Pudlox16}).
The two scenarios considered here correspond to scenarios 2 and 3 of the six scenarios detailed in the Supplementary Material Fig.~\ref{fig:Hum6}.
We chose to focus on these two scenarios 2 and 3 because they correspond to the scenarios that are both 
the most difficult to discriminate and the most compatible with the human SNP data available 
(see Section~\ref{sec:human_appli}).
For computational efficiency, our simulation-based study is limited to 5000 SNP genetic markers
(simulated with the Hudson criterion; see \citealt{Collin2021}) for 10 individuals sampled per population.
The parameters for both scenarios are
$\param = (N1, N2, N3, N4, N34, Na, t1, t2, t3, t4, Nbn3, Nbn4, Nbn34, d3, d4, d34, ra)$.
In both cases, the ancestral population has effective size (in number of diploid individuals) $Na$,
and becomes population 2 (ancestor to YRI) at time $t4$ (in number of generations since present), with effective size $N2$.
Population $34$ (ancestor to GBR and CHB) splits from population 2 at time $t3$,
and goes through a bottleneck with size $Nbn34$ for a duration $d34$,
before reaching a population size of $N34$.
Populations $3$ (ancestor to CHB) and $4$ (ancestor to GBR) then split at time $t2$,
and both undergo a bottleneck with respective sizes $Nbn3$ and $Nbn4$ for duration $d3$ and $d4$,
before reaching their respective population sizes $N3$ and $N4$.
Then at time $t1$, population 1 (ancestor to ASW and with effective size $N1$) is the result of an admixture event with population 4 (in scenario 2) or population 3 (in scenario 3) as parental populations , with an admixture rate $ra$ (proportion of genes with a non-African origin).
In these scenarios, we imposed the condition $t4>t3>t2$.
We used the same simulation program diyabc \citep{Collin2021}
and summary statistics as in the previous ``Dep-Indep''
case study.

We used the same priors for both models $m_0$ and $m_1$:
\begin{equation*}
    \param \sim \mathcal{P}
    \quad \iff \quad
    \left\{
   \begin{aligned}
    Ni & \sim \mathcal{U}(10^3,10^4) \text{ for } i \in \set{1, 2, 3, 4, 34};\\
    Na &\sim \mathcal{U}(10^2,10^3);\\
    t1 &\sim \mathcal{U}(1,30);\\
    ti &\sim \mathcal{U}(10^2,10^3) \text{ for } i \in \set{2, 3, 4};\\
    Nbni & \sim \mathcal{U}(5,500) \text{ for } i \in \set{3, 4, 34};\\
    di & \sim \mathcal{U}(5,500) \text{ for } i \in \set{3, 4, 34};\\
    ra &\sim \mathcal{U}(0.05,0.95),
    \end{aligned}
    \right.
\end{equation*}
and defined the two Bayesian models:
\begin{equation*}
   \text{$m_0$ (Sc. 2): }
      \left\{
   \begin{aligned}
   \param &\sim \mathcal{P}; \\
    \rawObs &\sim \mbox{Scenario\ 2}(\param);\\
    \yobs &= \text{SumStat}(\rawObs; m);
    \end{aligned}
    \right.
   \quad
   \text{$m_1$ (Sc. 3): }
   \left\{
   \begin{aligned}
   \param &\sim \mathcal{P}; \\
    \rawObs &\sim \mbox{Scenario\ 3}(\param);\\
    \yobs &= \text{SumStat}(\rawObs; m).
    \end{aligned}
    \right.
\label{eq:human_models}
\end{equation*}

\subsubsection{Power Computations}

For a given level of test $\alpha$, we define the power $\Pi$ of a \gof test for model $m_0$
by the probability of its p-value to be smaller than $\alpha$ when applied to a model that is
different from $H_0$ under the alternative model $m_1$. Power computations for pre- and post-inference \gof 
are detailed in the Supplementary Material Section~\ref{sec:power}, where the number of pseudo-observed datasets (POD) used a test datasets is also discussed.

\subsubsection{Test Calibration}

To test for the calibration of both tests, we reproduced the analyses described in the Supplementary Material Section~\ref{sec:power}, but this time using \PODs from the same prior predictive $\priorPredModel{m_0}$
used for the reference table, and checked whether the estimated p-values
from the $\ntest$ \PODs approximately followed a uniform distribution.

\subsubsection{Score Computation} \label{sec:score_computation_choice_k}
We computed the max-\lof score over $k$ varying from $5$ to $20$,
following standard recommendations \citep{Breunig2000}.
For the \knn, we used $k=1$ (nearest neighbor only), which we found to empirically give better results.
Results for other values of $k$ are presented in the Supplementary Material 
Sections~\ref{sec:prior_gof_all_k} and~\ref{sec:hpc_gof_all_K}.

\subsubsection{Practical Implementation}
All the scores and \gof tests are implemented in the \texttt{R} package
\texttt{abcgof}, freely available at \url{https://github.com/pbastide/abcgof}.
We implemented critical functions such as distance computations using
\texttt{C++} and package \texttt{Rcpp} \citep{Eddelbuettel2011} for improved speed,
and parallelized calculations using package \texttt{doParallel} \citep{doParallel}.

To compute p-values, both the \knn and \lof scores require the computation of the $k$
distances to the nearest neighbors of the test and calibration points 
to the reference points, for a varying value of $k$.
A naive implementation can compute all distances between calibration 
and reference points, and then sort the distances.
Here, we used the \knn implementation from the \texttt{dbscan} \texttt{R}
package, that relies on a k-d tree search to speed up the computation of
the $k$ nearest neighbors \citet{Hahsler2019}.

All the code to reproduce the figures and analyses of the paper
is available on GitHub: \url{https://github.com/pbastide/gof_sbi_paper}.

\subsection{Results}
The main results of our simulation study for the three settings presented above 
are described here. Further analyzes of these simulations are
detailed in Supplementary Material Section~\ref{sec:sim}.

\subsubsection{\lof is a good metric for \gof tests}
\begin{figure}[H]
    \centering
    \includegraphics[scale=0.8]{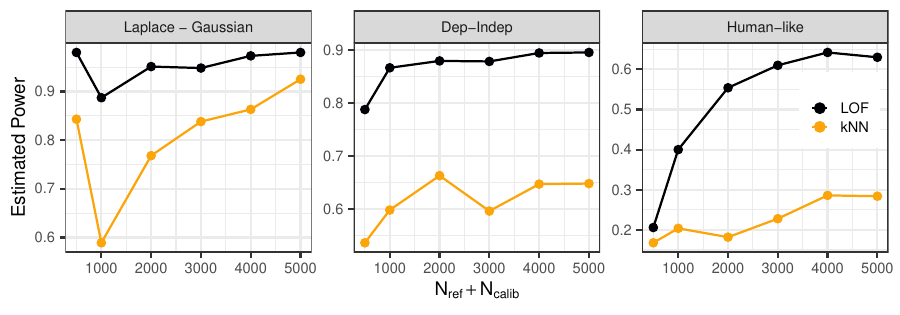}
    \caption{
    \textbf{Estimated power for the pre-inference (prior) \gof test}, at a level of $5\%$,
    for the three simulation settings (Laplace-Gaussian, Dep-Indep and Human-Like).
    The \lof score (black) is the ``max-\lof'' score for $k$ between $5$ and $20$.
    The \knn score (light orange) is computed for $k=1$.
    In all settings, $\nref = \ncalib$, and the total number of simulated particules is
    $\nref + \ncalib$ (x axis).
    }
    \label{fig:prior_GoF}
\end{figure}

Compared to the previous study of \citet{Lemaire2016}, we chose here to use the
\lof score instead of the \knn in all p-values computations, with the idea that,
the \lof score being specifically adapted for outlier detection, it will be
better at discriminating data points that are not from the null distribution.
In all of our tests, we saw that the \lof score indeed performed better than the
\knn, with higher power (see Fig.\ \ref{fig:prior_GoF} and \ref{fig:hpc_GoF}),
and better calibration 
(see Fig.\ \ref{fig:hpc_GoF_pval}).

\subsubsection{Pre-inference \gof is Powerful and Well Calibrated}
In particular, the pre-inference (prior) \gof test associated with the \lof metric had very good power,
above or close to $0.9$ even for a small number of simulated points in the two
``Laplace-Gaussian'' and ``Dep-Indep'' simple simulation settings (Fig.\ \ref{fig:prior_GoF}).
In the more complex ``Human-like'' setting, which includes two closely related scenarios, More simulated particles were needed to achieve a reasonable power of about $0.65$ with the \lof score, which is much higher than the about $0.3$ obtained with the \knn score.
As expected, the pre-inference \gof also proved to be well calibrated, with p-values under the null
hypothesis being approximately uniformly distributed
(see Supplementary Material Figure~\ref{fig:prior_gof_pval}).
We also tested a localized version of the pre-inference \gof, as in \citet{Lemaire2016}.
This test did not exhibit a significative increase in power, despite
being much more computationally intensive
(see Supplementary Material Figure~\ref{fig:prior_gof_localized}).

\subsubsection{Post-inference \gof Power is Robust to the Posterior Approximation}
\begin{figure}[H]
    \centering
    \includegraphics[scale=0.8]{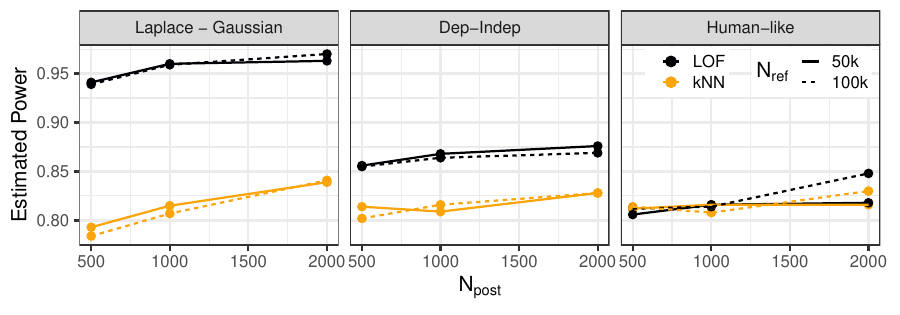}
    \caption{
    \textbf{Estimated power for the post-inference holdout \gof test},
    at a level of $5\%$,
    for the three simulation settings (Laplace-Gaussian, Dep-Indep and Human-like)
    and the rejection ABC posterior estimation method.
    The posterior was localized with $\epsilon$ such that the number of posterior particles
    is equal to $\npost$ (x axis),
    starting from $\nref$ = $50\;000$ (full line) and $100\;000$ (dashed line) total number of simulated particles. This corresponds to $0.5\%$ to $4\%$ of the simulations, depending on $\nref$.
    The \lof score (black) is the ``max-\lof'' score for $k$ between $5$ and $20$.
    The \knn score (light orange) is computed for $k=1$.
    In all settings, and $\ncalib'=\nref'=\npost/2$.
    }
    \label{fig:hpc_GoF}
\end{figure}

The post-inference \gof test had good power, i.e. above $0.8$,
even in the most complex ``Human-Like'' simulation setting (Fig.\ \ref{fig:hpc_GoF}).
The posterior inference method had little impact on the computed power, the 
simple rejection method performing as well or better than the local linear or ridge regression
(see Supplementary Figure~\ref{fig:hpc_GoF_supp}).
Doubling the number of particles, from $50$ to $100$ thousands, or changing the level of 
localization, from $0.5\%$ to $4\%$, had little effect.

\subsubsection{Post-inference \gof Calibration is Sensitive to the Posterior Approximation}
\begin{figure}[H]
    \centering
    \includegraphics[scale=0.8]{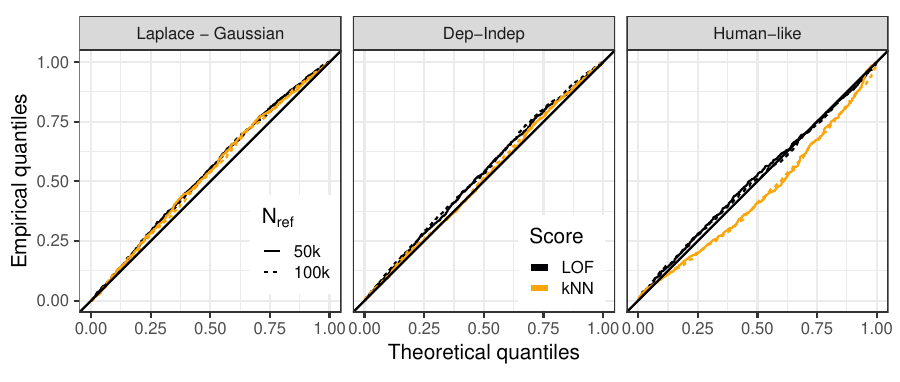}
    \caption{
    \textbf{Distribution of p-values of the post-inference \gof test}
    when the \PODs were simulated according to the null model, 
    for the three simulation settings (Laplace-Gaussian, Dep-Indep and Human-like)
    and the rejection ABC posterior estimation method.
    The total number $\nref$ of simulated particles was taken to be equal to 
    $50\;000$ (full lines) or $100\;000$ (dashed lines),
    and $1\;000$ points were taken in the posterior sample, with $\ncalib'=\nref'=\npost/2$.
    The \lof score (black) is the ``max-\lof'' score for $k$ between $5$ and $20$.
    The \knn score (light orange) is computed for $k=1$.
    }
    \label{fig:hpc_GoF_pval}
\end{figure}

The simple Rejection method with the \lof score seemed to be relatively robust, with
the empirical quantiles of the p-values close to the 
theoretical uniform quantiles (Fig.\ \ref{fig:hpc_GoF_pval}),
yielding for approximately calibrated tests.
As expected, the quality of the posterior approximation had however a substantial effect on the
calibration of the post-inference \gof test (see Supplementary Figure \ref{fig:hpc_GoF_pval_supp}).
In the simple ``Laplace - Gaussian'' setting, that has only two parameters, 
the local linear regression seems enough to correct the posterior estimation, and to
provide calibrated p-values.
In the ``Dep-Indep'' setting, the ridge regression seems to be needed
to correctly accommodate for the greater number of parameters.
Finally, in the more complex ``Human Like'' setting, none of the simple posterior correction
measures seemed to provide for calibrated p-values.

\section{Illustration using a real SNP dataset of modern Human populations}\label{sec:human_appli}

\subsection{Material and Methods}

\subsubsection{Data}

To illustrate our methodological developments, we analyzed a real SNP dataset obtained 
from individuals originating from four Human populations 
(30 unrelated individuals sequenced per population) using the freely accessible public 
dataset from
\citet{McVean2012}\footnote{vcf format files including variant calls are available at \url{http://www.1000genomes.org/data}.}.
The four Human populations included the Yoruba population (Nigeria) as representative of Africa (encoded YRI in the 1000 genome database), 
the Han Chinese population (China) as representative of the East Asia (encoded CHB), 
the British population (England and Scotland) as representative of Europe (encoded GBR), 
and the population composed of Americans of African Ancestry in SW-USA (encoded ASW).
The SNP loci were selected from the $22$ autosomal chromosomes using the following criteria:
(i) all $30\times4$ analyzed individuals have a genotype characterized by a quality score 
$\mbox{(GQ)}>10$ (on a PHRED scale), 
(ii) polymorphism is present in at least one of the $30\times4$ individuals in order to fit the SNP simulation algorithm used in DIYABC Random Forest v1.0 \citep{Collin2021}, 
(iii) the minimum distance between two consecutive SNPs is $1$ kb in order to minimize linkage disequilibrium between SNP, and
(iv) SNP loci showing significant deviation from Hardy-Weinberg equilibrium at a $1\%$ threshold \citep{Wigginton2005} in at least one of the four populations has been removed ($35$ SNP loci concerned).
After applying the above criteria, we obtained a dataset including $51\,250$ SNP loci scattered over the $22$ autosomes (with a median distance between two consecutive SNPs equal to $7$ kb),
among which two subsets each of $12\,000$ SNP loci
(thereafter referred to as dataset 1 and dataset 2) were randomly chosen for applying our \gof methods.

\subsubsection{Evolutionary Scenarios}

We considered six scenarios (i.e.\ models) of evolution of the four Human populations which differ from each other by one ancient and one recent historical event:
(i) A single out-of-Africa colonization event giving an ancestral out-of-Africa population which secondarily split into one European and one East Asian populational lineage, versus two independent out-of-Africa colonization events, one giving the European lineage and the other one giving the East Asian lineage. The possibility of a second ancient (i.e.\ older than $100\,000$ years) out of Africa colonization event through the Arabian peninsula toward Southern Asia has been suggested by archaeological studies (e.g., \citealp{Rose2011}).
(ii) The possibility (or not) of a recent genetic admixture of the Americans of African Ancestry in SW-USA between their African ancestors and individuals of European or East Asia origins (e.g., \citealp{Bryc2015}).
We detail the six considered evolutionary scenarios in the Supplementary Material Figure~\ref{fig:Hum6}.
Human population history has been and is still studied in details in a number of studies using genetic data (e.g., \citealp{Lohmueller2021}). Our aim here is to illustrate the potential of our pre-inference and post-inference \gof methodologies using real data on a well known case-study that exhibits complex evolutionary histories.

\subsubsection{Simulation of SNP Datasets and Summary Statistics}
We used the same parameters and priors for simulations as in Section~\ref{sec:sim_human_like},
with conditions on time events such that $t4>t3>t2$ for scenarios 1, 2 and 3,
and $t4>t3$ and $t4>t2$ for scenarios 4, 5 and 6.
We used the same diyabc and DIYABC-RF simulation tools \citealp{Collin2021}
and the same 130 summary statistics
as in Section~\ref{sec:depindep_setting}.

\subsubsection{Pre-inference \gof for Scenario Pruning}

We used the pre-inference (prior) \gof method to quickly, i.e.\ with a low computational cost, eliminate from the six proposed scenarios those characterized by a low propensity to generate multidimensional particles with summary statistics close to the particle corresponding to the observed dataset 1 (or the observed dataset 2).
To this aim, we simulated a reference table including $500$ to $5\,000$ particles 
(i.e. SNP datasets summarized by the above mentioned summary statistics) per scenario.
We then used the pre-inference \gof module of our new \texttt{abcgof} \texttt{R} package to compute the probability values of each scenario using the \knn outlier score (with $k=1$) or the \lof outlier score (using the ``max-\lof'' score for $k$ between $5$ and $20$); 
see Section~\ref{sec:methods} for details.
The probability values of each scenario were calculated for a number of simulations equal to $500$ (from which $250$ particles were used for calibration), 
$1\,000$ ($500$ for calibration),
$2\,000$ ($1\,000$ for calibration),
$3\,000$ ($1\,500$ for calibration),
$4\,000$ ($2\,000$ for calibration),
and $5\,000$ ($2\,500$ for calibration). 
$95\%$ confidence intervals were computed based on asymptotic standard error estimation
or on $500$ bootstrap replicates, as detailed in Section~\ref{sec:bootstrap} below.
To avoid inflated Type I error rates due to multiple testing issues, we applied the Benjamini-Hochberg \citep{Benjamini1995} correction to the bootstrap upper bounds of the \lof p-values computed with $2000$ particles.

\subsubsection{Scenario Choice Using ABC Random Forest}

Following \citet{Pudlox16}, we used Random Forests to select the best scenario between the two scenarios that were not eliminated after the pre-inference \gof treatment (i.e. scenarios 2 and 3).
We simulated a training set (i.e. reference table) including $10\,000$ particles for each of the scenarios 2 and 3 using the same set of 130 statistics plus one LDA axis (i.e., the number of scenarios minus 1 as predictors \citep{Pudlox16}.
We then used the ``Random Forest analyses'' module of DIYABC Random Forest v1.0 \citep{Collin2021}
to process model choice on the training set to predict the best scenario and estimate its posterior probability. 
We fixed the number of trees in the constructed Random Forests to $1\,000$, as this number turned out to be large enough to ensure a stable estimation of the global error rate (results not shown).

\subsubsection{Post-Inference \gof Test on the Selected Scenario}

We applied the post-inference \gof test on the above-selected scenario (i.e. scenario 2) to evaluate whether the selected scenario missed some important evolutionary features to explain the observed dataset, so that the multi-parameter posterior distribution conditional on the observed dataset has a poor propensity to generate multi-dimensional summary statistics particles close to the particle corresponding to the observed dataset. 
To this aim, we simulated a reference table including $50\,000$ or $100\,000$ particles under scenario 2. 
The target (i.e. the particle of the observed dataset) was either the particle of the observed dataset 1, with the particle of the observed dataset 2 used as replicate, or the particle of the observed dataset 2, with the particle of the observed dataset 1 used as replicate.
Two overlapping criteria were used to implement our post-inference \gof calculation.
The first criterion was to obtain a sufficiently strong localization around the particle corresponding to the observed dataset by retaining only the $\epsilon$ simulated particles, with $\epsilon < 5\%$, that are closest. 
The second criterion was to have a sufficiently large number of retained particles (more than $500$) to allow a reasonably robust linear or ridge regression in the local region defined above, as well as a reasonably robust calibration using half, and thus more than $250$, of the retained particles for calibration (the other $250$ being used to calculate probabilities);
see Section~\ref{sec:methods} for details.
We thus computed post-inference \gof probabilities using $\epsilon$ equal $1\%$, $2\%$ and $4\%$ for the reference table with $50\,000$ particles or $\epsilon$ equal $0.5\%$, $1\%$ and $2\%$ for the reference table with $100\,000$ particles.
We used the post-inference \gof module of our new \texttt{abcgof} \texttt{R} package to compute the post-inference probability value of scenario 2 using the rejection method or the rejection followed by a regression (linear or ridge) method, with the \knn score (for $k=1$) or the \lof score (fusing the ``max-\lof'' score for $k$ between $5$ and $20$).
As explained in Section~\ref{sec:post_gof}, the \gof post-inference method necessitates to have simulation replicate of each particle included in the subset of the $\epsilon$ retained particles.
We used the \texttt{–o} option of the command-line simulation software diyabc v1.1.51 to simulate replicates of the $\epsilon$ retained particles from their associated parameter vector values.
$95\%$ confidence intervals were computed based on asymptotic standard error estimation
or on $500$ bootstrap replicates, as detailed in Section~\ref{sec:bootstrap} below.

\subsection{{Quantification of uncertainty on p-values}}\label{sec:bootstrap}
In the pre-inference \gof case, the estimator of Equation~\eqref{eq:prior_GoF_emp} is unbiased
($\Esp[\hat{P}^{\text{prior}}(\yobs)] = P^{\text{prior}}(\yobs)$),
and has a variance
$\Var[\hat{P}^{\text{prior}}(\yobs)] = P^{\text{prior}}(\yobs) (1 - P^{\text{prior}}(\yobs))/\ncalib$
that decreases linearly with $\ncalib$.

Under the assumptions for good posterior estimation, the same is true for the post-inference \gof case.
This formula gives us a computationally cheap way of computing 
asymptotic confidence intervals, using a simple plug-in estimator for the standard error,
from a single draw of calibration points among the reference table.

A different approach, more costly in terms of computing time, to uncertainty assessment is to use a bootstrap procedure.
In the prior \gof test, given a certain number of simulations, the first step
is to choose which points we use as ``calibration'' points, and which we
use as ``reference'' points, which we do by simply sampling uniformly without
replacements points to be in the calibration set.
We can repeat this sampling several times, in order to get different sets
of calibration points from the simulated points, on which we apply the
procedure to compute the p-values.
In the post-inference \gof test, the same procedure can be applied after
the localization step, among the selected or inferred points from the
posterior.
In both case, this gives us a distribution of p-values, that we can use to get a point 
estimate (here using the median) and a confidence interval, here using 
the highest density interval (HDI, \citealp{Hyndman1996,Kruschke2015})
at a $95\%$ level.

\subsection{Results}

\subsubsection{Pre-inference \gof for Scenario Pruning}

\begin{figure}[H]
    \centering
    \includegraphics[scale=0.8]{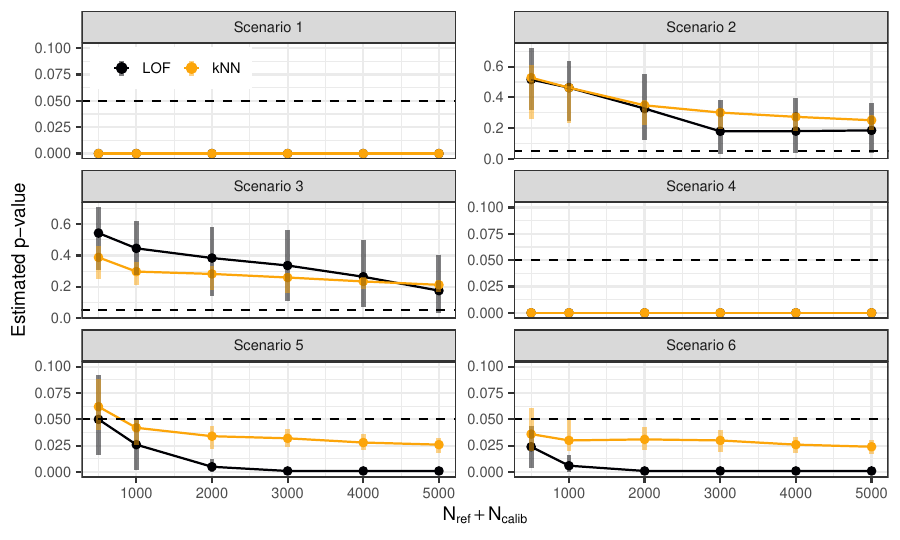}
    \caption{
    \textbf{Pre-inference \gof results on the real SNP dataset 1 of modern Human populations for
    different simulation efforts using the \lof or the \knn outlier score.}
    Estimated p-values for the prior \gof test, for the six scenarios of Figure~\ref{fig:Hum6}.
    The \lof score (black) is the ``max-\lof'' score for $k$ between $5$ and $20$.
    The \knn score (light orange) is computed for $k=1$.
    In all settings, $\nref = \ncalib$, and the total number of simulated particles is
    $\nref + \ncalib$ (x axis).
    Vertical bars represent the $95\%$ Highest Density Intervals (HDI) computed from
    the bootstrap procedure described in Section~\ref{sec:bootstrap} with $500$ replicates.
    See Supplementary Material Figure~\ref{fig:human_prior_gof_table_full}
    for a comparison with asymptotic intervals.
    The $5\%$ rejection threshold is shown as a dashed line.
    Only scenarios 2 and 3 are consistently not rejected,
    even at low computational cost (i.e.\ $1000$ simulations per scenario). The black and orange curves are superimposed for scenarios 1 and 4, so that only the orange curve is visible for these scenarios.
    }
    \label{fig:human_prior_GoF_table}
\end{figure}

Prior-\gof probability values were high (i.e. larger than $5\%$) for only two of the six scenarios considered, i.e. for scenarios 2 and 3
(see Figure~\ref{fig:human_prior_GoF_table} for results for the SNP dataset 1).
This result was well supported for both the \lof and \knn outlier scores, with upper bounds of 
the $95\%$ confidence intervals (using the bootstrap method) larger than $5\%$, 
when the number of simulations was larger than $500$ particles per scenario.
For a number of simulations of $500$ particles per scenario, the probabilities of scenario 5 were close to $5\%$ (i.e. $0.05$ and $0.06$ for the \lof and \knn scores, respectively), 
and the upper bounds of the $95\%$ intervals became close to $10\%$ for this scenario.
We note that the $95\%$ confidence intervals were significantly narrower when calculated 
using the asymptotic rather than the bootstrap method
(see Supplementary Material Figure~\ref{fig:human_prior_gof_table_full}).
Consistent with our simulation-based studies, which show a higher power for the \lof score than for the \knn score, the pre-inference \gof probability values tend to be larger for the \knn than for the \lof score.
When applying the Benjamini-Hochberg \citep{Benjamini1995} correction to the bootstrap upper bounds of the \lof p-values with $\nref+\ncalib=2000$, we still rejected
all scenarios but scenarios 2 and 3 at the $5\%$ level.
We obtained similar results for the real SNP dataset 2 (results not shown).

As a reminder, the purpose of the scenario pruning step is to limit the simulation of unnecessary data from clearly ``out-of-game'' models due to the high computational cost involved.
We show here on real data that even for complex scenarios with a large number of parameters, this goal can be achieved with a limited number of simulations on the order of $1\,000$ particles per scenario. Overall, our pre-inference \gof results show that the scenarios 1, 4, 5 and 6 can be set aside, leaving only the scenarios 2 and 3 to be kept for further simulations and inference.
The next inference step is to apply statistical treatments of model choice to select the best between scenarios 2 and 3.
To this end, we simulated a total of $10\,000$ particles for each of the scenarios 2 and 3 and used random forests for the selection of the best scenario. Scenario 2 was clearly the best, with posterior probabilities of $1.000$ and $0.999$ for the real datasets 1 and 2, respectively.

\subsubsection{Post-inference \gof on the selected scenario}

\begin{figure}[H]
    \centering
    \includegraphics[scale=0.8]{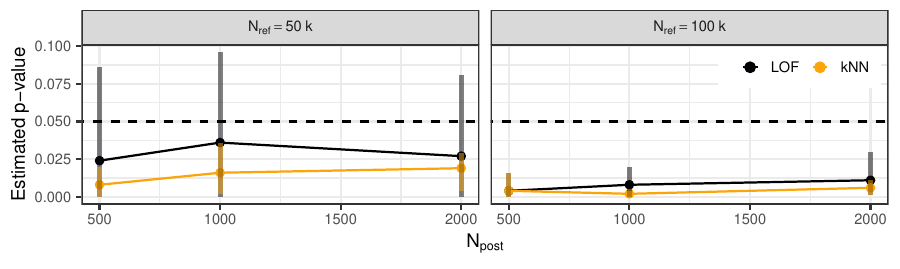}
    \caption{
        \textbf{Post-inference \gof results on the real SNP dataset 1 of modern Human populations for different simulation efforts using the \lof or the \knn outlier score.}
        Estimated p-values for the post-inference \gof test for the 
        selected scenario 2 of Figure~\ref{fig:Hum6},
        using the rejection ABC posterior estimation method.
        The posterior was localized with $\epsilon$ such that the number of posterior particles
        is equal to $\npost$ (x axis),
        starting from $50\;000$ (left panel) and $100\;000$ (right panel) total number of simulated particles.
        In all settings, $\ncalib'=\nref'=\npost/2$.
        The \lof score (black) is the ``max-\lof'' score for $k$ between $5$ and $20$.
        The \knn score (light orange) is computed for $k=1$.
        Vertical bars represent the $95\%$ Highest Density Intervals (HDI) computed from
        the bootstrap procedure described in Section~\ref{sec:bootstrap} with $500$ replicates.
        The $5\%$ rejection threshold is shown as a dashed line.
        Tested scenarios 2 gets consistently rejected by the post-inference test.
        See Supplementary Material Figures~\ref{fig:human_post_gof_table_bootstrap}
        and~\ref{fig:human_post_gof_table_asymptotic}
        for a comparison with other posterior estimation methods and for asymptotic intervals.
    }
    \label{fig:human_post_GoF_table}
\end{figure}

For the rejection method, post-\gof probability values of scenario 2 were small for both the \lof and \knn scores, with upper bounds of the $95\%$ CIs below $5\%$, indicating that scenario 2 poorly fits the observed datasets 
(see Figure~\ref{fig:human_post_GoF_table} for results for the SNP dataset 1).
This result was slightly less well supported when the number of simulations was $50\,000$ (compared to $100\,000$), as indicated by upper bounds of the $95\%$ HDI that were above $5\%$ (but below $10\%$) for the \lof score (but not for the \knn score). 
Our simulation-based studies on the human-like case have shown a similar power for the \lof score than for the \knn score. In agreement with this,  the post-inference \gof probability values obtained on real data were similar for both outlier scores, but $95\%$ CIs were narrower for the \knn than the \lof scores.
We note again that, as for the pre-inference \gof, the $95\%$ confidence intervals were significantly narrower when calculated from a single draw of calibration particles than when calculated from $500$ bootstrapped iterations of calibration particles
(see Supplementary Material Figure~\ref{fig:human_post_gof_table_bootstrap} and~\ref{fig:human_post_gof_table_asymptotic}).

For the local linear and ridge regression methods, the post-inference \gof probabilities of scenario 2 are higher than for the rejection method 
(see Supplementary Material Figure~\ref{fig:human_post_gof_table_bootstrap}). Probability values of scenario 2 using the the local linear and ridge regression methods show strong variations depending on the number of simulations and $\epsilon$ parameter, and generally do not indicate that scenario 2 fits the observed data poorly (contrary to what was found using the rejection method).
Overall, these results are consistent with our simulation-based studies on the human-like case, which show a higher power for the rejection method than for the local linear (but not for the ridge regression) methods, as well as a clearly better calibration for the rejection method.

For both with or without-regression post-inference \gof methods, we observe a substantial decrease in the probability values and their variation when we move from $50\,000$ to $100\,000$ simulations, as well as when $\epsilon$ values are getting smaller (Figure~\ref{fig:human_post_GoF_table} and Supplementary Material Figure~\ref{fig:human_post_gof_table_bootstrap}). 
However, even in such more ``optimal'' conditions, the evidence in favour of the rejection of the scenario 2 remain weak and uncertain for the two  post-inference \gof methods using a regression step.
For instance for $100\,000$ simulations and $\epsilon = 0.005$, the post-inference \gof probability of the method using local linear regression is $0.072$ ($95\%$CI $[0.036, 0.104]$) for \knn and $0.056$ ($95\%$CI $[0.004; 0.216]$) for \lof. We obtained similar results for the real dataset 2 (results not shown).

\section{Discussion}

In population genetics as well as many other application fields, inferences about models with intractable likelihoods rely on simulation-based methods, such as Approximate Bayesian Computation and Simulation-Based Inference, for tasks such as parameter inference and model selection. Once a model has been selected, the assessment of its Goodness of Fit (\gof) presents a significant challenge, particularly in the absence of a likelihood function. We introduce two novel \gof tests based on Local Outlier Factor (LOF) that more effectively captures local density deviations than a more traditional k-Nearest Neighbors approach. The pre-inference (prior) \gof test serves to prune the less plausible models (with a low computational cost), while the post-inference \gof test evaluates the fit of the selected model in greater detail. This dual approach provides a comprehensive assessment of model validity, combining Bayesian and general goodness-of-fit insights.
The dual-test \gof approach employed by our methodology underscores its flexibility. The pre-inference \gof test offers insight into model validity from a Bayesian perspective, while the post-inference test provides a more general and traditional view of assessing goodness of fit. 

\subsection{Biological Insights for the Human Population Genetics Dataset}
Our application of the pre-inference prior \gof methodology for modern human evolutionary scenario pruning demonstrates that it is feasible to avoid the need for extensive, data-intensive simulations for four apparently implausible scenarios among the six complex scenarios examined. This goal can be effectively achieved with a modest simulation volume, i.e.  around 1\;000 particles per scenario. The four scenarios identified as unlikely include events that are inconsistent with current knowledge of the evolution of \textit{Homo sapiens}. Specifically, they posit two independent colonization events outside Africa, one leading to European populations and the other to northeast Asian populations (e.g. Chinese populations distinct from those of South Asia along coastal regions; \cite{Lopez2015}), or they overlook genetic admixture among American individuals of African descent \citep{Bryc2015}.

The two scenarios that continue to be supported by high probabilities of pre-inference \gof are closely related in terms of genetic history, as they both propose a single colonization event outside Africa and an admixture source for African American ancestry, either from Europe (scenario 2) or Asia (scenario 3). These two regions contain genetically weakly differentiated populations, making it difficult to distinguish conclusively between scenarios 2 and 3 at this stage. A more definitive decision in favor of scenario 2 requires a formal scenario choice test, which we perform here using a random forest based classification on 10\;000 simulations generated specifically for scenarios 2 and 3. This formal approach provides a more reliable means of distinguishing between the scenarios, given their nuanced genetic similarities.

Applying safely our post-inference \gof methodology to scenario 2 requires a higher number of simulations for the selected scenario only, of the order of $100\;000$. The post-\gof treatments, at least those based on the rejection method, show that (even) scenario 2 does not give a good account of the real observed Human datasets (i.e. low post-inference \gof probability well below $5\%$). This result is not surprising as a number of human population evolutionary modalities highlighted in the last two decades are not modeled in scenario 2. These include more complex demographic fluctuations than those modeled in scenario 2, the mixing of ancestral European and Asian populations with Neanderthal and Denisovan human lineages respectively, several successive waves of colonization of the European area, and recurrent episodes of back into Africa migrations (e.g. reviewed in \citealp{Lopez2015}).

\subsection{Calibrated Post-Inference \gof Requires Good Posterior Estimation}
The accuracy of the post-inference likelihood test is contingent on the quality of the posterior approximation. While traditional methods, such as rejection-based Approximate Bayesian Computation (ABC), offer a straightforward approach to posterior approximation, they frequently exhibit deficiencies related to bias and inefficiency. These limitations necessitate the employment of more sophisticated posterior approximation techniques. Recent advancements in machine learning have introduced powerful methods for posterior approximation, such as normalizing flows, specifically Masked Autoregressive Flows (MAF, \cite{PapamakariosPM17}), and Distributional Random Forests (DRF, \cite{CevidMNBM22}). MAF offer a flexible way to approximate complex posterior distributions by learning dependencies between parameters in a tractable manner. Sequential Neural Likelihood \citep{PapamakariosSM19}, a technique that has emerged as a promising method for refining posterior approximations through iterative learning, thereby improving the convergence and accuracy of the inference, takes advantage of MAF. Furthermore, importance sampling and adaptive ABC methods present additional opportunities to enhance the posterior by refining simulations in regions of higher posterior density. Also, the use of a non-parametric approach, like the DRF, can adapt to the structure of the data while mitigating the risk of overfitting.

\subsection{More Useful Models}

Finally, as George Box famously stated, ``All models are wrong, but some are useful''. This perspective is particularly relevant in the context of Bayesian generative models, where the goal is not to find a perfect representation of reality but rather to develop a model that is useful for making inferences and predictions. Given the inherent simplifications and assumptions made in modeling, it is crucial to interpret the results of \gof tests with an understanding of their limitations. Model misspecification is an unavoidable aspect of statistical inference. While \gof tests are essential for identifying discrepancies between observed data and model predictions, they can be viewed as tools for iterative model refinement rather than definitive tests of validity. When a model fails a \gof test, several actionable steps can be considered: (i) refinement of priors and summary statistics: adjusting priors to better capture known biological constraints or selecting more appropriate summary statistics can significantly improve model fit; and (ii) incorporating additional complexity: introducing features such as population structure, environmental influences, or time-varying parameters can enhance model realism.
The iterative process of model criticism and refinement is key to advancing our understanding of complex systems, and \gof tests can serve as a critical component of this process.

\section*{Scripts and Data}

Data and scripts to reproduce the analyses presented in this paper.
Also available on GitHub: \url{https://github.com/pbastide/gof_sbi_paper}

\section*{Acknowledgments}

We thank Paul Verdu for useful discussions about recent findings regarding Human populations evolution. 
We are grateful to the INRAE MIGALE bioinformatics facility 
(MIGALE, INRAE, 2020. Migale bioinformatics Facility, doi: 10.15454/1.5572390655343293E12)
for providing computing resources.

\section*{Funding}

This work has been supported by funds from the project 
AgroStat (reference: ANR-23-EXMA-0002)
of the Maths-VivES France 2030 program 
handled by the French \textit{Agence Nationale de la Recherche}.

\newpage 

\appendix

\begin{center}
\Huge \textbf{Supplementary Material}
\end{center}

\renewcommand{\thefigure}{S\arabic{figure}}
\setcounter{figure}{0}

\hspace{1cm} \section{Localized Prior \gof}\label{sec:local_prior_gof}

\subsection{Localized Prior \gof}

\citet{Lemaire2016} developed a localized version of the prior \gof,
that tests the prior null hypothesis of Equation~\eqref{eq:prior_GoF_hyp},
but, instead of using the simple reference table $\refTable$,
use a localized version of it around the observation $\yobs$.
The estimated p-value for this test is then defined as:
\begin{equation}\label{eq:loc_prior_GoF_emp}
\hat{P}^{\text{loc prior}}(\yobs) =
\frac{1}{\ncalib} \sum_{\icalib = 1}^{\ncalib} 
\ind{\scoreRefTable{\y_{\icalib}}{\refTableHat^{\lvert\yobs}} > \scoreRefTable{\yobs}{\refTableHat^{\lvert\yobs}}},
\end{equation}
where $\refTableHat^{\lvert\yobs}$ is obtained as in
Equation~\eqref{eq:holdout_GoF_emp} of the main text,
by re-simulating the lines of the reference tables that are the closest
to the observation (in a simple rejection approach).
This test hence has the same computational cost as the post-inference \hpc \gof.

Note that is test is not a posterior predictive check, as the null
hypothesis corresponds to the prior predictive distribution,
which is why we refer to it as a localized prior \gof test.

\subsection{Performance on simulations}

We used this localized prior \gof on the same three simulation settings as
in the main text, simulating $50$ or $100$ thousand points for each model,
and using a rejection localized with $\epsilon = 1\%$.
We compared the results of the localized prior \gof to the results of the
standard prior \gof described in Section~\ref{sec:prior_gof} of the main text.

\begin{figure}[H]
    \centering
    \includegraphics[scale=0.8]{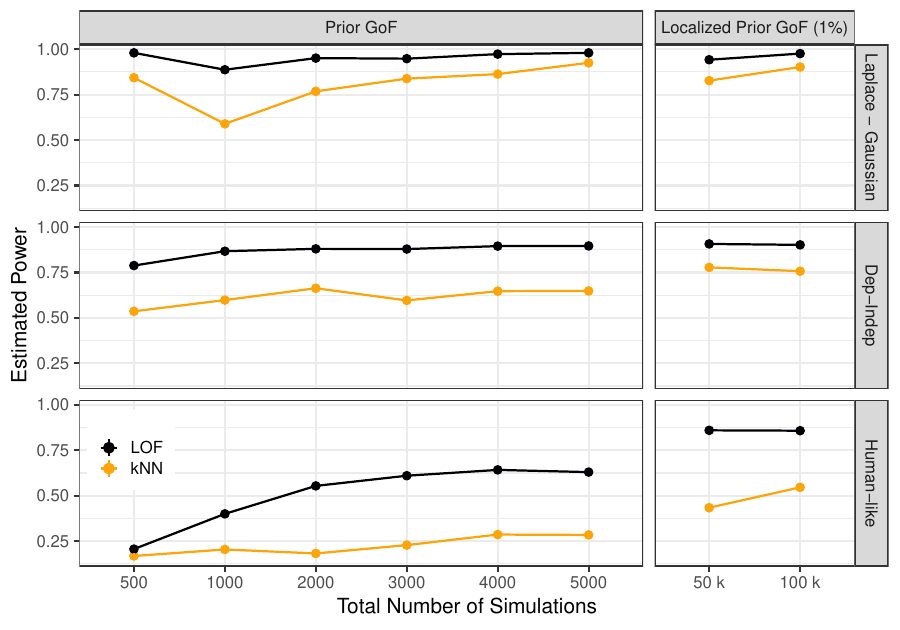}
    \caption{
    \textbf{Estimated power for the prior (left) and localized prior (right) \gof tests},
    at a level of $5\%$,
    for the three simulation settings.
    The \lof score (black) is the ``max-\lof'' score for $k$ between $5$ and $20$.
    The \knn score (light orange) is computed for $k=1$.
    In all settings, $\nref = \ncalib$, and the total number of simulated particules is
    $\nref + \ncalib$ (x axis).
    }
    \label{fig:prior_gof_localized}
\end{figure}

Figure~\ref{fig:prior_gof_localized} shows the compared power of the two approaches.
We can see that, despites being much more computationally intensive
(it requires $10$ to $200$ times more simulations from the model),
the localized \gof does not improve the power at all in the
Laplace-Gaussian and Dep-Indep examples
(Fig.\ \ref{fig:prior_gof_localized}, first two lines).
In the more complex Human-like example, the power increases only slightly.

We hence advocate for using the light weight prior \gof as described in the main text,
and, if computational power is available, to turn to the holdout \gof instead
of the localized prior \gof, that does not exhibit clear power gains.

\section{Power Computations}\label{sec:power}

For a given level of test $\alpha$, we define the power $\Pi$ of a \gof test for model $m_0$
by the probability of its p-value to be smaller than $\alpha$ when applied to a model that is
different from $H_0$ under the alternative model $m_1$.

\subsection{Pre-inference \gof}

For the pre-inference (prior) \gof of Equations~\ref{eq:prior_GoF_def}, we get:
\begin{align*}
\Pi^{\text{prior}}
&= \Proba_{\yobs \sim \priorPredModel{m_1}}
\left[
P^{\text{prior}}(\yobs) \leq \alpha
\right]
\\
&= \Proba_{\yobs \sim \priorPredModel{m_1}}
\left[
\Proba_{\yrep \sim \priorPredModel{m_0}}\left[\scoreModel{\yobs}{m_0} > \scoreModel{\yrep}{m_0} \right] 
\leq \alpha
\right].
\label{eq:prior_power_def}
\end{align*}
To estimate this quantity, we draw $\ntest$ observations from the prior predictive of
alternative model $m_1$: $\y^1_{\itest} \underset{\text{iid}}{\sim} \priorPredModel{m_1}$, $1 \leq k \leq \ntest$,
and we approximate both probabilities by their frequencies as in Equation~\ref{eq:prior_GoF_emp},
assuming that we have calibration points from the prior predictive of $m_0$:
$\y^0_{\icalib} \underset{\text{iid}}{\sim} \priorPredModel{m_0}$, $1 \leq i \leq \ncalib$:
\begin{align*}
\hat{\Pi}^{\text{prior}} 
&= \frac{1}{\ntest} \sum_{\itest = 1}^{\ntest} 
\ind{\hat{P}^{\text{prior}}(\y^1_{\itest}) \leq \alpha}
\\
&= 
\frac{1}{\ntest} \sum_{\itest = 1}^{\ntest}
\ind{
\frac{1}{\ncalib} \sum_{\icalib = 1}^{\ncalib} \ind{
\scoreModel{\y^0_{\icalib}}{m_0} > \scoreModel{\y^1_{\itest}}{m_0}
}
\leq \alpha
}
\\
&=
\frac{1}{\ntest}
\card{\set{
\itest\ |\ \scoreModel{\y^1_{\itest}}{m_0} \geq \scoreModel{\cdot}{m_0}_{1-\alpha},\ 1\leq \itest \leq \ntest
}},
\end{align*}
with $\scoreModel{\cdot}{m_0}_{1-\alpha}$ the $1 - \alpha$ quantile of 
$\set{\scoreModel{\y^0_{\icalib}}{m_0} \ | \ 1 \leq \icalib \leq \ncalib}$,
and $\card{\cdot}$ the cardinal of the set.

For the computation of pre-inference \gof power, we took $\nref$ between $250$ and $2500$ for the reference table,
and used $\ncalib = \nref$ in Formula~\ref{eq:prior_GoF_emp}, so that the total number of simulations from $m_0$ varied from $500$ to $5000$.

\subsection{Post-inference \gof}

For the power of the post-inference (holdout) \gof test, we use the same definitions for the post-inference \gof, using 
Equations~\ref{eq:holdout_GoF_def} and~\ref{eq:holdout_GoF_emp}. We first draw $\y^1_{\itest} \underset{\text{iid}}{\sim} \priorPredModel{m_1}$, $1 \leq k \leq \ntest$,
observations from the prior predictive of model $m_1$. Then, for each $\itest$, we draw a replicate $\y^{1,\text{new}}_{\itest}$ using the same model $m_1$ on the parameters $\param_{\itest}$ that
were used for $\y^{1}_{\itest}$, so that $\y^{1,\text{new}}_{\itest}$
and $\y^{1}_{\itest}$ are drawn from the same model, with the same parameter
values.
The estimated power in this case is given by:
\begin{align*}
\hat{\Pi}^{\text{holdout}} 
&= \frac{1}{\ntest} \sum_{\itest = 1}^{\ntest} 
\ind{\hat{P}^{\text{holdout}}(\y^1_{\itest}, \y^{1,\text{new}}_{\itest}) \leq \alpha}
\\
&= 
\frac{1}{\ntest} \sum_{\itest = 1}^{\ntest}
\ind{
\frac{1}{\ncalib'} \sum_{\icalib = 1}^{\ncalib'} 
\ind{\scoreRefTable{\y^0_\icalib}{\refTableHatModel{m_0}^{\lvert\y^1_{\itest}}} >
\scoreRefTable{\y^{1,\text{new}}_{\itest}}{\refTableHatModel{m_0}^{\lvert\y^1_{\itest}}}}
\leq \alpha
}.
\end{align*}

For the computation of post-\gof power, we took $\nref \in \set{50\;000, 100\;000}$, and,
for each pseudo-observed dataset (\POD) $\y_{\itest}$, $1 \leq \itest \leq \ntest$, we estimated the posterior distribution
using a simple rejection-based ABC (see, e.g., \citealp{Marin2012})
by selecting the lines of the original reference table $\refTableModel{m_0}$
that were within a distance of $\epsilon$ of $\y_{\itest}$:
\begin{equation*}
    \refTableModel{m_0}^{\lvert\y^1_{\itest},\epsilon}
    =
    \set{d \in \refTableModel{m_0} | \|d - \y_{\itest}\| \leq \epsilon}.
\end{equation*}
We chose $\epsilon$ so as to keep a total of $\npost$ particles in the localized
dataset, with $\npost$ varying between $500$ and $2000$.
This corresponds to $0.5\%$ to $4\%$ of the simulations, depending on $\nref$.
For each line $\iref$ of $\refTableModel{m_0}^{\lvert\y^1_{\itest},\epsilon}$,
we then extracted the parameters $\param_i$ from the reference table,
and re-simulated a set of observations $\y_\iref$ using the same model
$m_0$ and the same parameters $\param_i$,
to get $\refTableHatModel{m_0}^{\lvert\y^1_{\itest},\epsilon}$.
This re-simulation aimed at approximating draws from the unknown posterior 
$\postPred{\cdot}{\y_\itest}$.

It is clear from Equation~\eqref{eq:holdout_GoF_emp} that the posterior approximation
method should have a strong impact on the calibration of the post-inference
\gof test. In addition to the simple Rejection ABC, we tested the local linear
regression with an Epanechnikov kernel \citep{Beaumont2002}, taking the median predictions as parameter point estimates, and the ridge regression, taking the median predictions from
three penalisation parameter values $\lambda$ equal to $0.0001$, $0.001$, $0.01$ \citep{Blum2010}.
In both cases, we selected the same lines $\refTableModel{m_0}^{\lvert\y^1_{\itest},\epsilon}$
from the reference table, but then computed $\hat{\param}_i$
using the regression, and finally re-simulated $\refTableHatModel{m_0}^{\lvert\y^1_{\itest},\epsilon}$
using these estimated parameters.

In all cases, we used again half of the lines for the calibration, that we picked at random
in the posterior reference table $\refTableHatModel{m_0}^{\lvert\y^1_{\itest},\epsilon}$,
so that $\ncalib' = \npost / 2$ and $\refTableHatModel{m_0}^{\lvert\y^1_{\itest}}$
had $\nref' = \npost / 2$ lines in Formula~\ref{eq:holdout_GoF_emp}.

\subsection{Number of test \PODs}

We took $\ntest = 1000$ for the first two simulation settings 
(i.e. the Toy ``Laplace-Gaussia'' toy Models and ``Dep - Indep'' population genetics scenarios), 
and $\ntest = 500$ for the third one (i.e. the ``Human-like'' evolutionary scenarios),
 which was more computationally demanding.

\section{Additional Simulation Results}\label{sec:sim}

\subsection{Parameter Transform}

When using a regression approach (local linear or ridge), the correction
term does not take the constraints of the parameters into account,
and can produce posterior predictions that lie outside of the set boundaries.
Instead of just rejecting these prediction, an other approach is to 
apply the regression correction in a transformed, unconstrained parameter space
\citep{Beaumont2010}.

For the ``Laplace-Gaussian'' and ``Dep-Indep'' examples, only independent
bound constraints were applied on parameters, so that we used independent
logistic transforms on each parameter, as in \citep{Blum2010}.
As the bounds are inclusive, to avoid infinite values, we used slightly inflated
bounds, by removing $0.49$ to the lower bound, and adding $0.49$ to the upper bound.
As all parameters are assumed to be integers for simulations,
we then rounded the obtained back-transformed values, which ensured that all values
lay between the bounds (bounds included).

For the ``Human-like'' and empirical Human datasets, 
in addition to bound constraints,
relative order constraints were applied on splitting times, with $t4 > t3 > t2$.
We dealt with this constraint using the logistic transformation of differences.
More specifically, setting all the constraints together, we have:
\[
t2 < t3 < t4 ; \;
100 \leq t2 \leq 998 ; \;
101 \leq t3 \leq 999 ; \;
103 \leq t4 \leq 1000.
\]
The transform we applied is then:
\[
\left\{
\begin{aligned}
u2 &= t2 \\
u3 &= \max(t3 - t2 - 1; \alpha) \\
u4 &= \max(t4 - t3 - 1; \alpha),
\end{aligned}
\right.
\quad
\text{and}
\quad
\left\{
\begin{aligned}
v2 &= \logit(u2; 100, 998) \\
v3 &= \logit(u3; 0, 998 - t2 + 0.49) \\
v4 &= \logit(u4; 0, 999 - t3 + 0.49),
\end{aligned}
\right.
\]
where $\alpha = 0.25$ is chosen so that the lower bound of $0$
on the difference is never attained, thus avoiding infinite values,
and $\logit(\cdot; a, b)$ is the logit (inverse logistic) transform
with upper and lower bounds $a$ and $b$.
This transformation sends the constrained parameter values into $\mathbb{R}^3$,
and can be inverted as:
\[
\left\{
\begin{aligned}
u2 &= \logitinv(v2; 100, 998) \\
u3 &= \logitinv(v3; 0, 998 - t2 + 0.49) \\
u4 &= \logitinv(v4; 0, 999 - t3 + 0.49),
\end{aligned}
\right.
\quad
\text{and}
\quad
\left\{
\begin{aligned}
t2 &= \round(u2) \\
t3 &= \round(u3 + t2 + 1) \\
t4 &= \round(u4 + t3 + 1),
\end{aligned}
\right.
\]
where the rounding operation ensure that we recover integers.
Note that this is not a one to one map, as we are mapping a finite set
(constrained integer parameters) into $\mathbb{R}^3$, and it is only injective.
Other approaches based on unconstrained parameters as in the ``Dep-Indep''
example could avoid the use of this ad-hoc transformation, that has
known issues \citep{Beaumont2010}.

\subsection{Impact of \texorpdfstring{$K$}{K} value on the prior \gof} \label{sec:prior_gof_all_k}

\begin{figure}[H]
    \centering
    \includegraphics[scale=0.7]{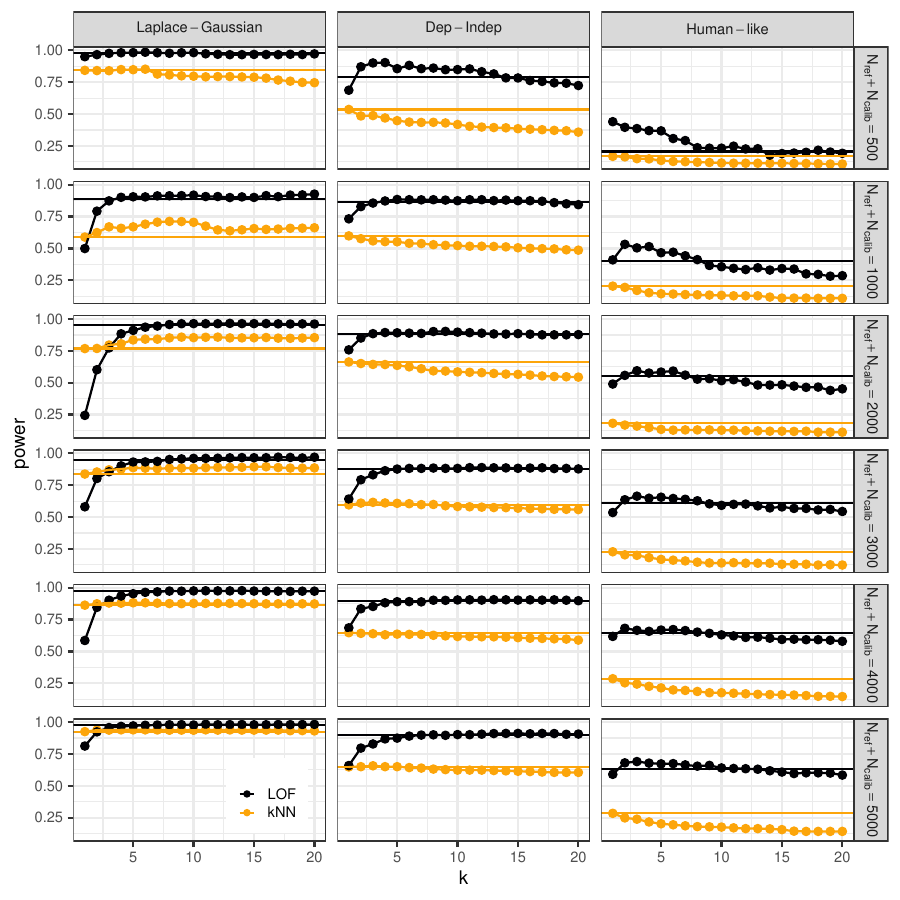}
    \caption{
    \textbf{Estimated power for the prior \gof test}, at a level of $5\%$,
    for the three simulation settings.
    The \lof score is computed for $k$ between $1$ and $20$ (black dots),
    and using the ``max-\lof'' score for $k$ between $5$ and $20$ (horizontal black line).
    The \knn score is computed for $k$ between $1$ and $20$ (light orange dots),
    and using a fixed $k=1$ (horizontal light orange line) as in the main text for comparison.
    In all settings, $\nref = \ncalib$, and the total number of simulated particles is
    $\nref + \ncalib$ (lines).
    }
    \label{fig:prior_gof_all_k}
\end{figure}

Figure~\ref{fig:prior_gof_all_k} shows the impact of the choice of $k$ in
the \lof and \knn scores on the power of the prior \gof test.
For the \knn score (light orange), the horizontal line corresponding to $k=1$
as in the main text almost always yields the highest power, especially in 
the more complex example (last column).
Similarly, for the \lof score (black), the horizontal line corresponding
to the ``max-\lof'' score for $k$ between $5$ and $20$
as in the main text allows for powers that are close to the optimal over
all $k$ values.
For both scores, the value of $k$ has a relatively small impact,
which justify the default values we used in the main text.

\subsection{Prior \gof p-value Calibration} \label{sec:prior_gof_pval}

\begin{figure}[H]
    \centering
    \includegraphics[scale=0.7]{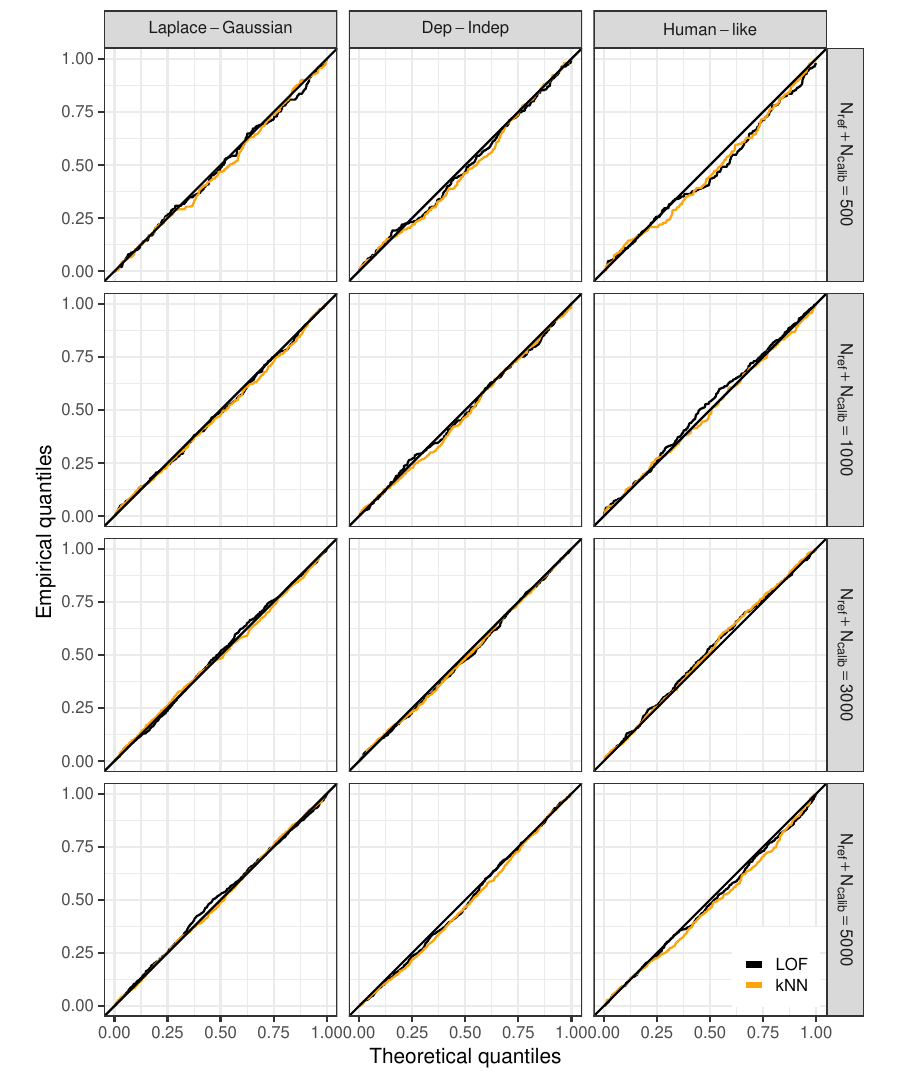}
    \caption{
    \textbf{Distribution of p-values of the post-inference \gof test}
    when the \PODs are simulated according to the null model, 
    for the three simulation settings 
    and the three ABC posterior estimation methods .
    The total number of simulated particles varies from $500$ to $5000$,
    with $\nref = \ncalib$ in all settings.
    The \lof score (black) is the ``max-\lof'' score for $k$ between $5$ and $20$.
    The \knn score (light orange) is computed for $k=1$.
    }
    \label{fig:prior_gof_pval}
\end{figure}

Figure~\ref{fig:prior_gof_pval} shows the distribution of p-values of the
prior \gof test when the \PODs are simulated from the null model.
It shows that the test is well calibrated, in all settings and for
both the \lof and \knn scores, with the p-values being approximately 
uniformly distributed.

\subsection{Impact of \texorpdfstring{$K$}{K} value on the Post Inference \gof Power} \label{sec:hpc_gof_all_K}
\begin{figure}[H]
    \centering
    \includegraphics[scale=0.7]{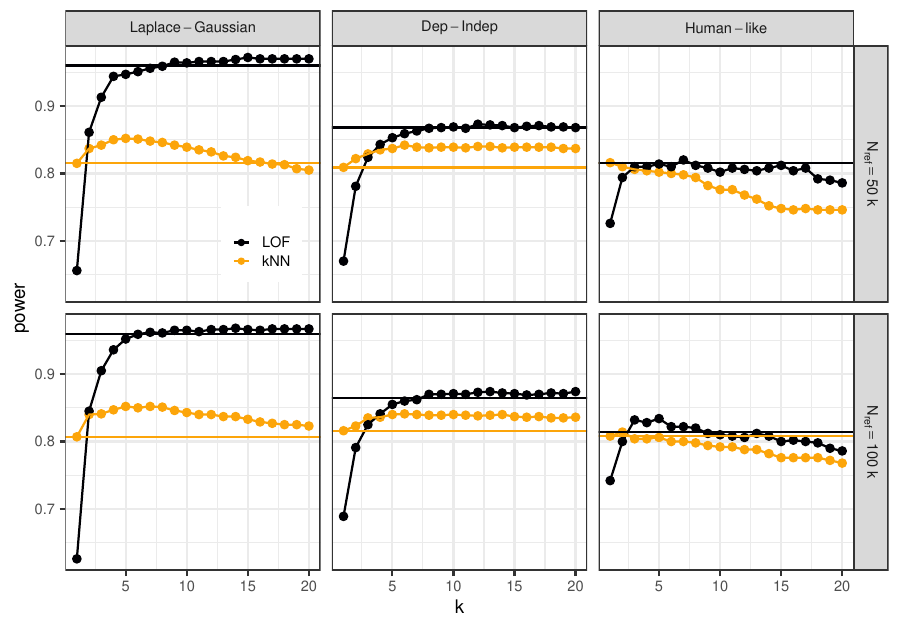}
    \caption{
    \textbf{Estimated power for the holdout \gof test}, at a level of $5\%$,
    for the three simulation settings.
    The posterior was localized with $\epsilon$ such that the number of posterior particles
    is equal to $\npost = 1000$,
    starting from $50\;000$ (top line) and $100\;000$ (bottom line) 
    total number of simulated particles.
    The \lof score is computed for $k$ between $1$ and $20$ (black dots),
    and using the ``max-\lof'' score for $k$ between $5$ and $20$ (horizontal black line).
    The \knn score is computed for $k$ between $1$ and $20$ (light orange dots),
    and using a fixed $k=1$ (horizontal light orange line) as in the main text for comparison.
    In all settings, $\nref' = \ncalib' = 500$.
    }
    \label{fig:hpc_gof_all_K}
\end{figure}

Figure~\ref{fig:hpc_gof_all_K} shows the impact of the choice of $k$ in
the \lof and \knn scores on the power of the holdout \gof test.
The results are similar to the ones we obtained for the prior \gof test,
with the default value of $k$ we chose having reasonable performances.

\subsection{Impact of the Posterior Estimation Method on the Post Inference \gof Power}
\begin{figure}[H]
    \centering
    \includegraphics[scale=0.7]{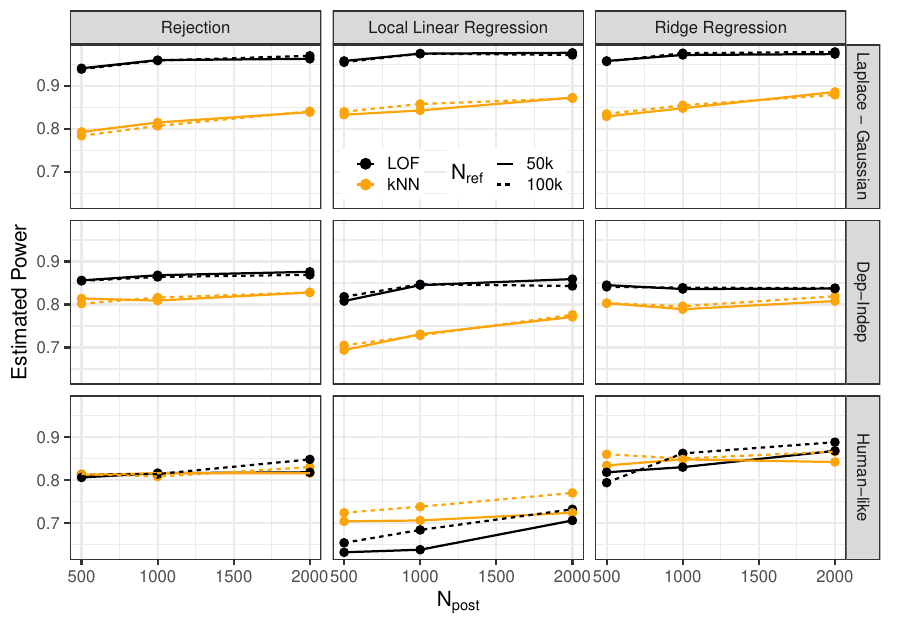}
    \caption{
    \textbf{Estimated power for the post-inference holdout \gof test},
    at a level of $5\%$,
    for the three simulation settings
    and the three ABC posterior estimation methods.
    The posterior was localized with $\epsilon$ such that the number of posterior particles
    is equal to $\npost$ (x axis),
    starting from $\nref$ = $50\;000$ (full line) and $100\;000$ (dashed line) total number of simulated particles. This corresponds to $0.5\%$ to $4\%$ of the simulations, depending on $\nref$.
    The \lof score (black) is the ``max-\lof'' score for $k$ between $5$ and $20$.
    The \knn score (light orange) is computed for $k=1$.
    In all settings, and $\ncalib'=\nref'=\npost/2$.
    }
    \label{fig:hpc_GoF_supp}
\end{figure}

Figure~\ref{fig:hpc_GoF_supp} shows the impact of the posterior estimation procedure on the
power of the holdout \gof test.
The three methods tested here yield similar powers.

\subsection{Impact of the Posterior Estimation Method on the Post Inference \gof Calibration}
\begin{figure}[H]
    \centering
    \includegraphics[scale=0.7]{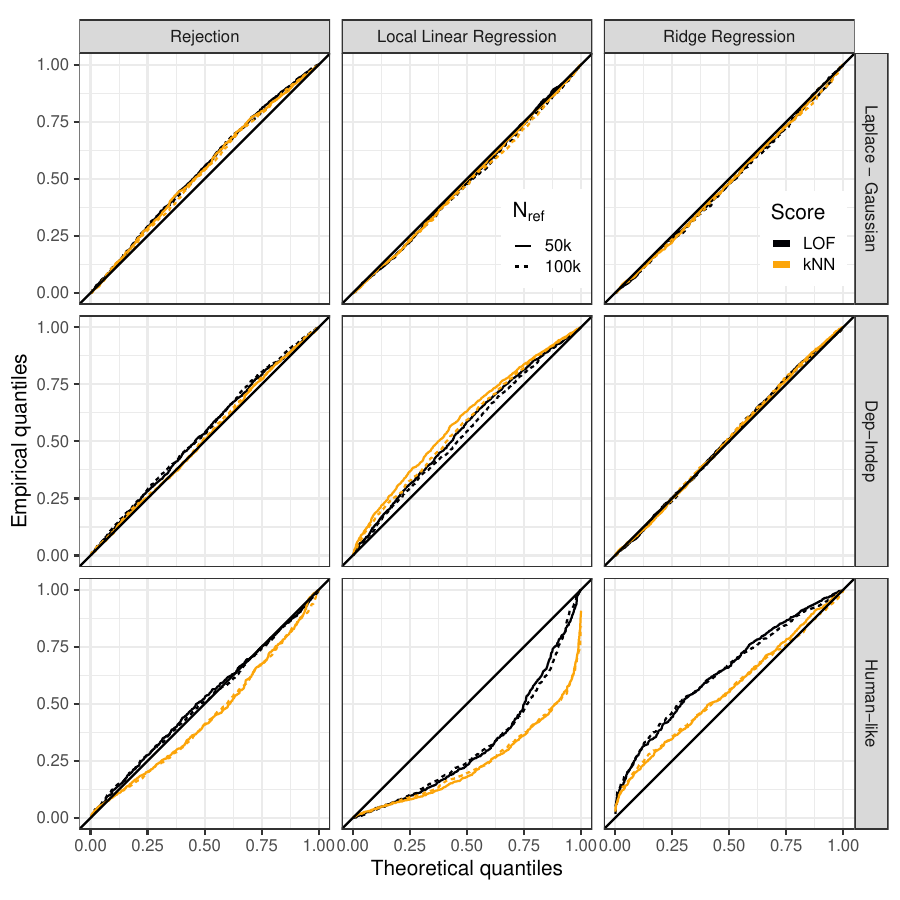}
    \caption{\textbf{Distribution of p-values of the post-inference \gof test}
    when the \PODs were simulated according to the null model, 
    for the three simulation settings (Laplace-Gaussian, Dep-Indep and Human-like)
    and the three ABC posterior estimation methods (rejection, local linear or ridge regression).
    The total number $\nref$ of simulated particles was taken to be equal to 
    $50\;000$ (full lines) or $100\;000$ (dashed lines),
    and $1\;000$ points were taken in the posterior sample, with $\ncalib'=\nref'=\npost/2$.
    The \lof score (black) is the ``max-\lof'' score for $k$ between $5$ and $20$.
    The \knn score (light orange) is computed for $k=1$.
    }
    \label{fig:hpc_GoF_pval_supp}
\end{figure}

Figure~\ref{fig:hpc_GoF_pval_supp} shows the strong impact of the posterior estimation procedure on the
calibration of the holdout \gof test.
While simple corrections seems to yield correctly calibrated tests in simple datasets
(first two rows), they do not seem to be sufficient in the complex example (last row).

\section{Human Population Genetics Example: The six Considered Evolutionary Scenarios} \label{sec:human_scenarios}

  \begin{figure}[H]
   \begin{center}
     \setlength{\unitlength}{5mm}
     \includegraphics[height=13cm,width=0.7\textwidth]{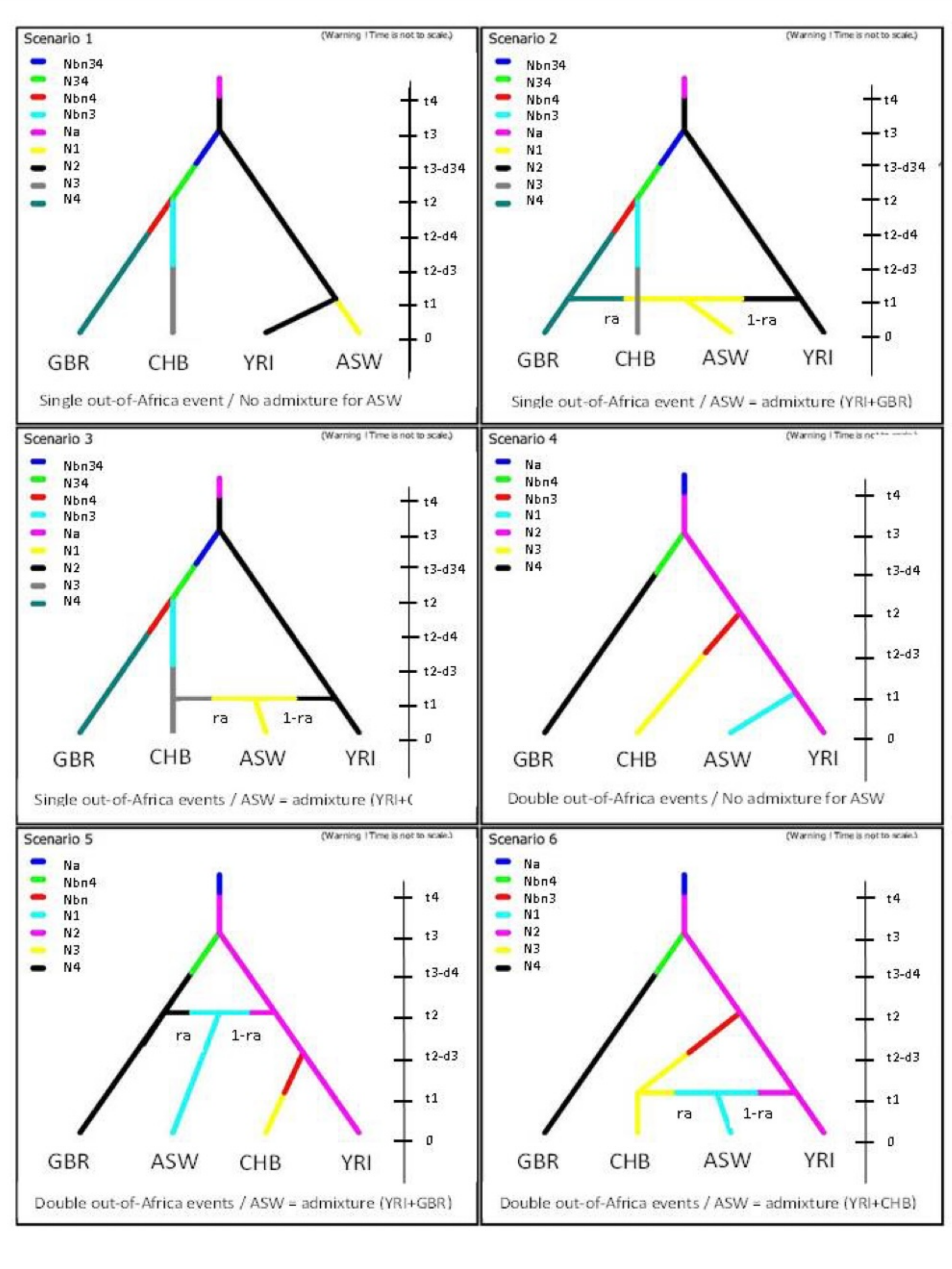}
   \end{center}
   \caption{
   \textbf{
   Six scenarios of evolution of four modern Human populations.}
   The genotyped populations are 
   Yoruba (YRI, Nigeria, Africa),
   Han (CHB, China, East Asia), 
   British (GBR, England and Scotland, Europe), 
   and Americans of African Ancestry (ASW, USA).
   The six scenarios differ from each other by one ancient and one recent historical event:
   (i) a single out-of-Africa colonization event giving an ancestral out-of-Africa population
   which secondarily split into one European and one East Asian population lineage
   (scenarios 1, 2 and 3),
   versus two independent out-of-Africa colonization events,
   one giving the European lineage and the other one giving the East Asian lineage
   (scenarios 4, 5 and 6).
   (ii) the possibility (or not; scenarios 1 and 4) of a recent genetic admixture of ASW individuals
   with their African ancestors and individuals of European (scenarios 2 and 5) or East Asia 
   (scenarios 3 and 6) origins.
   The scenarios 2 and 3 are those considered in the third case study of our power simulation-based study (see Section~\ref{sec:sim_human_like}). 
   }
   \label{fig:Hum6}
 \end{figure}

\section{Human Population Genetics Example: Additional Analyses}

\subsection{Human Population Genetics Example: Prior p-values}

\begin{figure}[H]
    \centering
    \includegraphics[scale=0.8]{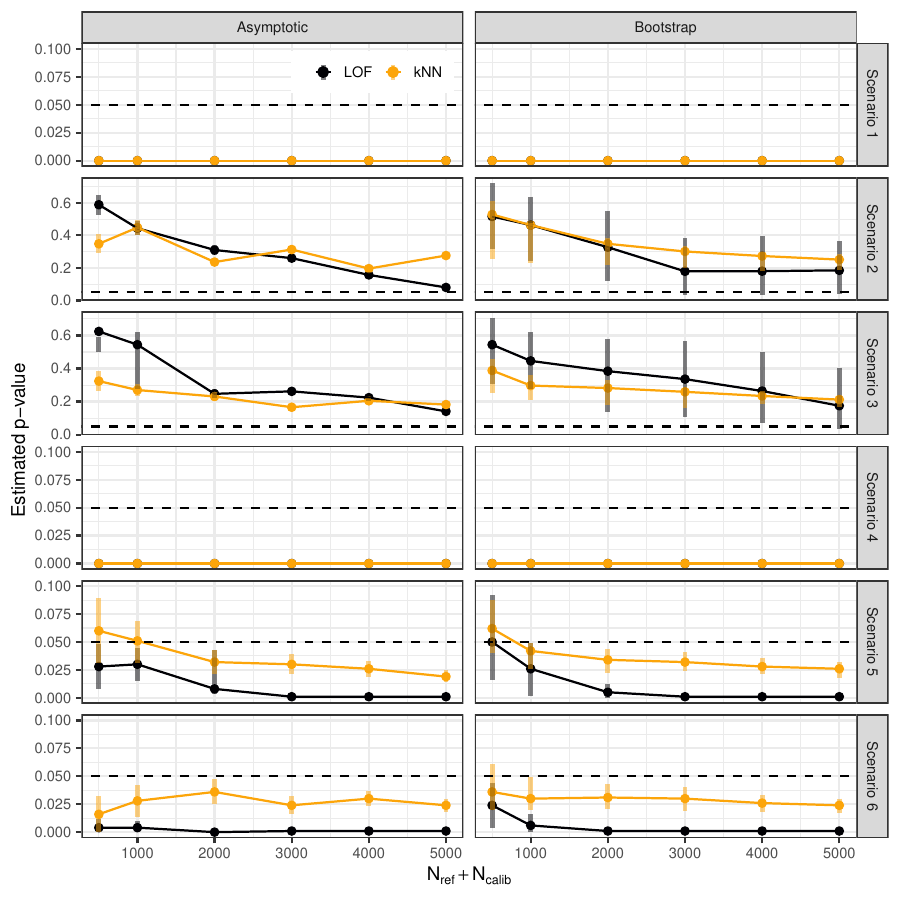}
    \caption{
    \textbf{Pre-inference GoF results on the real SNP dataset1 of modern Human populations for
    different simulation efforts using the LOF or the kNN outlier score.}
    Estimated p-values for the prior \gof test, for the six scenarios of Figure~\ref{fig:Hum6}.
    The \lof score (black) is the ``max-\lof'' score for $k$ between $5$ and $20$.
    The \knn score (light orange) is computed for $k=1$.
    In all settings, $\nref = \ncalib$, and the total number of simulated particles is
    $\nref + \ncalib$ (x axis).
    Vertical bars represent the $95\%$ confidence interval,
    either using the asymptotic formula (left column)
    or the Highest Density Intervals (HDI) computed from
    the bootstrap procedure with $500$ replicates (right column).
    The $5\%$ rejection threshold is shown as a dashed line.
    Only scenarios 2 and 3 are consistently not rejected,
    n particular for relatively low simulation costs
    (i.e.\ $1000$ simulations per scenario). 
    }
    \label{fig:human_prior_gof_table_full}
\end{figure}

Figure~\ref{fig:human_prior_gof_table_full} compares the confidence intervals
obtained with the asymptotic and bootstrap methods
(see Section~\ref{sec:bootstrap} of the main text).
Asymptotic intervals tend to be narrower than bootstrap intervals that,
although more computationally intensive, might better represent the
uncertainty around the point estimate of the p-value.

\subsection{Human Population Genetics Example: Post Inference p-values}
\begin{figure}[H]
    \centering
    \includegraphics[scale=0.8]{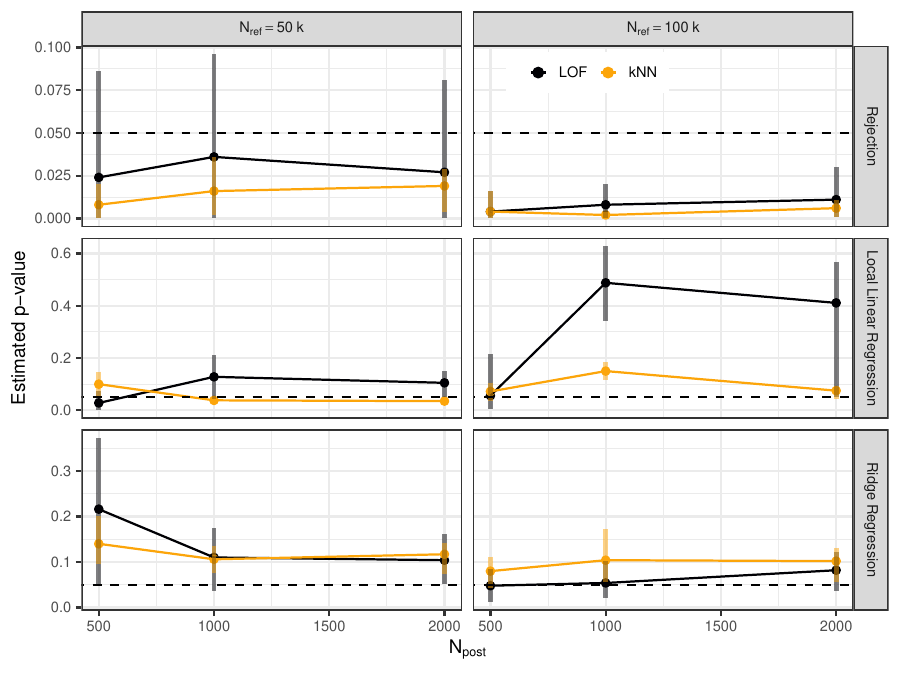}
    \caption{
        \textbf{Post-inference GoF results on the real SNP dataset1 of modern Human populations for different simulation efforts using the LOF or the kNN outlier score.}
        Estimated p-values for the holdout \gof test for the 
        selected scenario 2 of Figure~\ref{fig:Hum6},
        using the the three ABC posterior estimation methods
        (rejection, local linear or ridge regression).
        The posterior was localized with $\epsilon$ such that the number of posterior particles
        is equal to $\npost$ (x axis),
        starting from $50\;000$ (left panel) and $100\;000$ (right panel) total number of simulated particles.
        In all settings, $\ncalib'=\nref'=\npost/2$.
        The \lof score (black) is the ``max-\lof'' score for $k$ between $5$ and $20$.
        The \knn score (light orange) is computed for $k=1$.
        Vertical bars represent the $95\%$ Highest Density Intervals (HDI) computed from
        the bootstrap procedure described in Section~\ref{sec:bootstrap} with $500$ replicates.
        The $5\%$ rejection threshold is shown as a dashed line.
        }
        \label{fig:human_post_gof_table_bootstrap}
\end{figure}

\begin{figure}[H]
    \centering
    \includegraphics[scale=0.8]{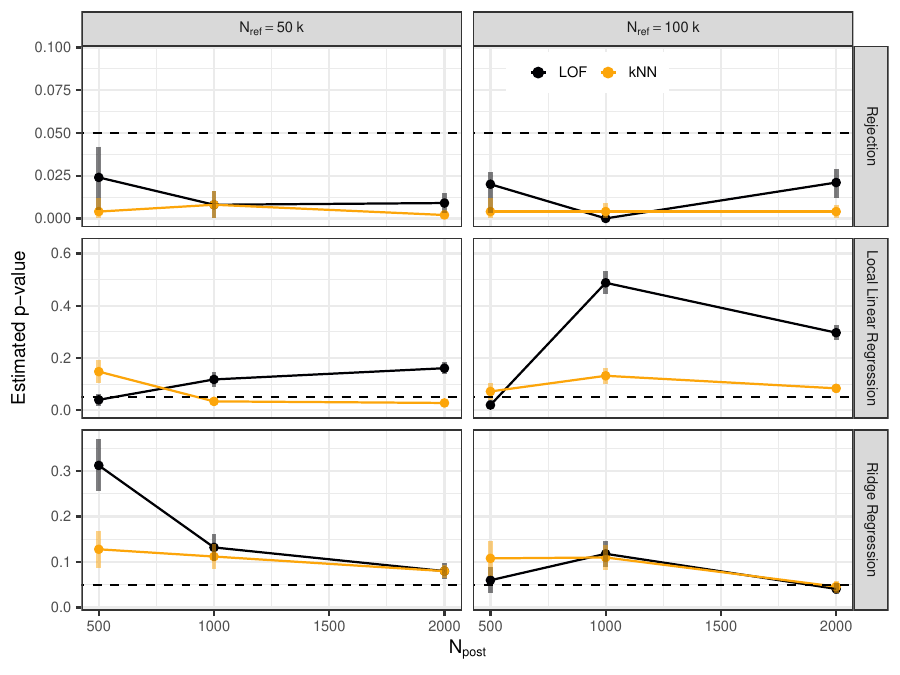}
    \caption{
        \textbf{Post-inference GoF results on the real SNP dataset1 of modern Human populations for different simulation efforts using the LOF or the kNN outlier score.}
        Estimated p-values for the holdout \gof test for the 
        selected scenario 2 of Figure~\ref{fig:Hum6},
        using the the three ABC posterior estimation methods
        (rejection, local linear or ridge regression).
        The posterior was localized with $\epsilon$ such that the number of posterior particles
        is equal to $\npost$ (x axis),
        starting from $50\;000$ (left panel) and $100\;000$ (right panel) total number of simulated particles.
        In all settings, $\ncalib'=\nref'=\npost/2$.
        The \lof score (black) is the ``max-\lof'' score for $k$ between $5$ and $20$.
        The \knn score (light orange) is computed for $k=1$.
        Vertical bars represent the $95\%$ Highest Density Intervals (HDI) computed from
        the asymptotic procedure described in Section~\ref{sec:bootstrap}.
        The $5\%$ rejection threshold is shown as a dashed line.
        }
    \label{fig:human_post_gof_table_asymptotic}
\end{figure}

Figures~\ref{fig:human_post_gof_table_bootstrap} and~\ref{fig:human_post_gof_table_asymptotic}
show the p-values estimated for the holdout test applied on Scenario 2,
using various posterior estimation methods,
and either the asymptotic or bootstrap uncertainty estimation procedure.
Although Scenario 2 gets consistently rejected when using the rejection method
(first line), the result is less clear for the linear and ridge regression procedures.
However, as shown in Figure~\ref{fig:hpc_GoF_pval} of the main text, 
using these posterior estimation methods tend to worsen the calibration of the 
test compared to the more simple rejection method, so that we are less confident in 
the results of these methods.
Better posterior estimation procedures would be needed to circumvent this issue.
As previously, asymptotic intervals tend to be narrower than bootstrap intervals
        
\newpage

\bibliographystyle{plainnat}
\bibliography{bibliography.bib}

\end{document}